\documentclass[12pt]{article}
\usepackage{amsmath}
\usepackage{graphicx,psfrag,epsf}
\usepackage{enumerate}
\usepackage{url} 

\usepackage{natbib}
\usepackage{xr}
\usepackage{xr-hyper}
\usepackage{hyperref}
\newcommand{\blind}{0}

\usepackage{enumitem}
\setlist{nosep} 

\usepackage{booktabs}
\usepackage{array}
\newcolumntype{M}{>{\centering\arraybackslash}m{\dimexpr.05\linewidth-2\tabcolsep}}

\usepackage[usenames,dvipsnames]{xcolor}

\graphicspath{{figures/}}

\usepackage{todonotes}

\usepackage{bm}
\usepackage{amstext}
\usepackage{amssymb}
\usepackage{amsmath}
\usepackage{amsfonts}
\usepackage{multirow}

\newcommand{\inv}{\ensuremath{^{-1}}}

\newcommand{\trans}{\ensuremath{^\prime}}

\begin{document}

\def\spacingset#1{\renewcommand{\baselinestretch}%
{#1}\small\normalsize} \spacingset{1}

\if0\blind
{
  \title{\bf  Model Choice and Diagnostics for Linear Mixed-Effects Models Using Statistics on Street Corners}
  \author{Adam Loy\thanks{
    The authors gratefully acknowledge funding from the National Science Foundation Grant \#DMS 1007697. All data collection has been conducted with approval from the Institutional Review Board IRB 10-347}\hspace{.2cm}\\
    Department of Mathematics, Lawrence University\\
    and \\
    Heike Hofmann\\
    Department of Statistics and Statistical Laboratory, Iowa State University\\
    and \\
    Dianne Cook\\
    Department of Econometrics and Business Statistics, Monash University
    }
  \maketitle
} \fi

\if1\blind
{
  \bigskip
  \bigskip
  \bigskip
  \begin{center}
    {\LARGE\bf  Model Choice and Diagnostics for Linear Mixed-Effects Models Using Statistics on Street Corners}
\end{center}
  \medskip
} \fi

%
%

\bigskip
\begin{abstract}
The complexity of linear mixed-effects (LME) models means that traditional diagnostics are rendered less effective. This is due to a breakdown of asymptotic results, boundary issues, and visible patterns in residual plots that are introduced by the model fitting process. Some of these issues are well known and adjustments
have been proposed. Working with LME models  typically requires that the analyst keeps track of all the special circumstances that may arise. In this paper we illustrate a simpler but generally applicable approach to diagnosing LME models. We explain how to use new visual inference methods for these purposes. The approach
provides a unified framework for diagnosing LME fits and for model selection. 
We illustrate the use of this approach on several commonly available data sets. 
A large-scale Amazon Turk study was used to validate the methods. R code is provided for the analyses.
\end{abstract}

{\it Keywords:} Statistical graphics; Lineup protocol; Visual inference; Model diagnostics;  Model selection
\vfill

\clearpage
\spacingset{1.45}
\section{Introduction}


Model checking is an essential step of statistical modeling that ensures the  assumptions necessary for valid inference are upheld. This process includes both the search for contradictions of model assumptions and an assessment of how well the model captures the characteristics of the data. Such investigation can be carried out using test statistics and $p$-values to gauge the strength of evidence, however, such methods indicate only the degree to which 
there is a problem with the model. Graphical diagnostics enable the analyst to detect not only when there is a problem with the model and where it occurs, but also give some indication of  
what may be the cause of the problem. For complex models, the insight provided by graphical diagnostics is especially useful, allowing the analyst to develop intuition and make discoveries about the nature of model violations. This would be nearly impossible through purely numeric methods. 

\begin{figure}[!h]
	\centerline{
	\includegraphics[width=0.22\textwidth]{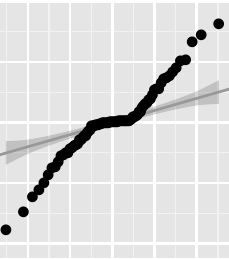} \hfill 
	\includegraphics[width=0.22\textwidth]{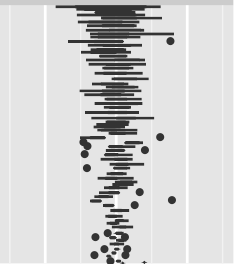} \hfill 
	\includegraphics[width=0.22\textwidth]{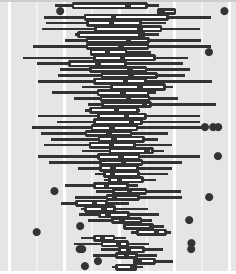} \hfill 
	\includegraphics[width=0.22\textwidth]{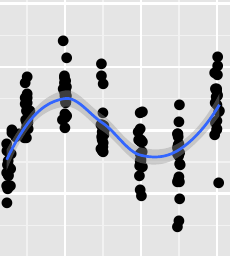}
	}
	\caption{\label{fig:motivation} Four residual plots from different LME models. Which ones, or all, of these would you consider to indicate structure not captured by the model?
	}
\end{figure}

While graphical diagnostics develop intuition and make discoveries, it is easy to misinterpret a single plot, especially when dealing with more complex models such as linear mixed effects (LME) models.
For example, Figure~\ref{fig:motivation} displays four residual plots used to diagnose LMEs: (left) a normal Q-Q plot of the random slopes; (center) box plots displaying the error terms by group; and (right) a plot of the residuals vs. a predictor. Which of the plots in Figure~\ref{fig:motivation} would you consider indicative of a model violation?

Based on our discussions with other statisticians, we have found that all four plots could be deemed indicative of a violation; however, only the plot of the residuals vs.\ the predictor is truly problematic.  Figure~\ref{fig:motivation} demonstrates that using only a single plot to distinguish between structure introduced by deviations from the model and structure introduced by the model fitting procedure is difficult, especially for more complicated statistical models such as LMEs. The purpose of this paper is to illustrate how visual inference allows us to distinguish between structure due to deviations from the model and structure introduced by the model fitting procedure. Additionally, the proposed graphical tests can be used in situations where the assumptions of conventional inferential procedures are violated. While we focus only on LMEs, the discussion is general enough to illustrate how visual inference can be used to overcome similar issues in other model classes. 

The remainder of this paper is organized as follows.  
Section~\ref{sec:vi} provides an introduction to the framework of visual inference and the lineup protocol. 
Section~\ref{sec:lme} introduces problems encountered in model selection and model checking, which are expanded upon in 
 Sections~\ref{sec:select} and \ref{sec:checking}, respectively, and presents solutions utilizing visual inference. 
Throughout Sections~\ref{sec:select} and \ref{sec:checking} multiple examples are used, and comparisons between conclusions on model building and fit from the new visual inference approach and existing diagnostic tools are provided. All example data sets are readily available for public use. A short description of each data set can be found in Section~\ref{supp:datasets} of the supplement.


%
%

\section{Visual inference}\label{sec:vi}

Classical statistical inference consists of 
(i) formulating  null and alternative hypotheses,
	(ii) calculating a test statistic from the observed data,
	(iii) comparing the test statistic to a reference (null) distribution,
	and (iv) deriving a $p$-value on which a conclusion is based.
Each of these steps has a direct analog in visual inference, as outlined by \cite{Buja:2009hp}. 
This section highlights the  parallels between conventional hypothesis tests and visual inference in the setting of LME models.

Assume that  the question of interest involves  some assumption about a model (such as a null hypothesis of homogeneity of residual variance) while the alternative hypothesis  encompasses any violation of this model assumption. 
For visual inference, the test statistic corresponds to a plot  that displays an aspect of  the model assumption and allows the observer to distinguish between scenarios under the null hypothesis from scenarios under alternative hypotheses. 
Plots drawn from data generated consistently with the null hypothesis are called \emph{null plots}. The set of all null plots constitutes the reference distribution; thus, the plot of the observed data is indistinguishable from the null plots if the model assumption holds.
In the lineup protocol, the plot of the observed data is randomly embedded
 among a sample of, usually 19, null plots drawn from the reference distribution.  These \emph{lineups} are then presented to independent observers for evaluation. 

Evaluation by independent observers  allows for the estimation of a $p$-value associated with the lineup:
 Let $X$ be the random variable describing  the number  of observers, out of $N$, identifying the data plot. 
If $X=x$ is  the number of observers who chose the data plot from the lineup, then the  $p$-value is  the probability that at least $x$ observers chose the data plot, given that the null hypothesis  is true (i.e.,~the  data plot is not any different from the other plots in the lineup). Under the null hypothesis the probability of choosing the true plot is $1/m$ (for a lineup of size $m$), and $X$ is distributed according to a distribution similar to a Binomial distribution $B_{N, 1/m}$, but adjusted for the dependencies between plots in a given lineup.
\citep[][introduced visual $p$-values. Details of the calculation for this LME model application are given in Section~\ref{sec:pvalues} of the supplement.]{mahbub:2013}

\begin{figure}
	\centering
	\includegraphics[width=0.8\textwidth]{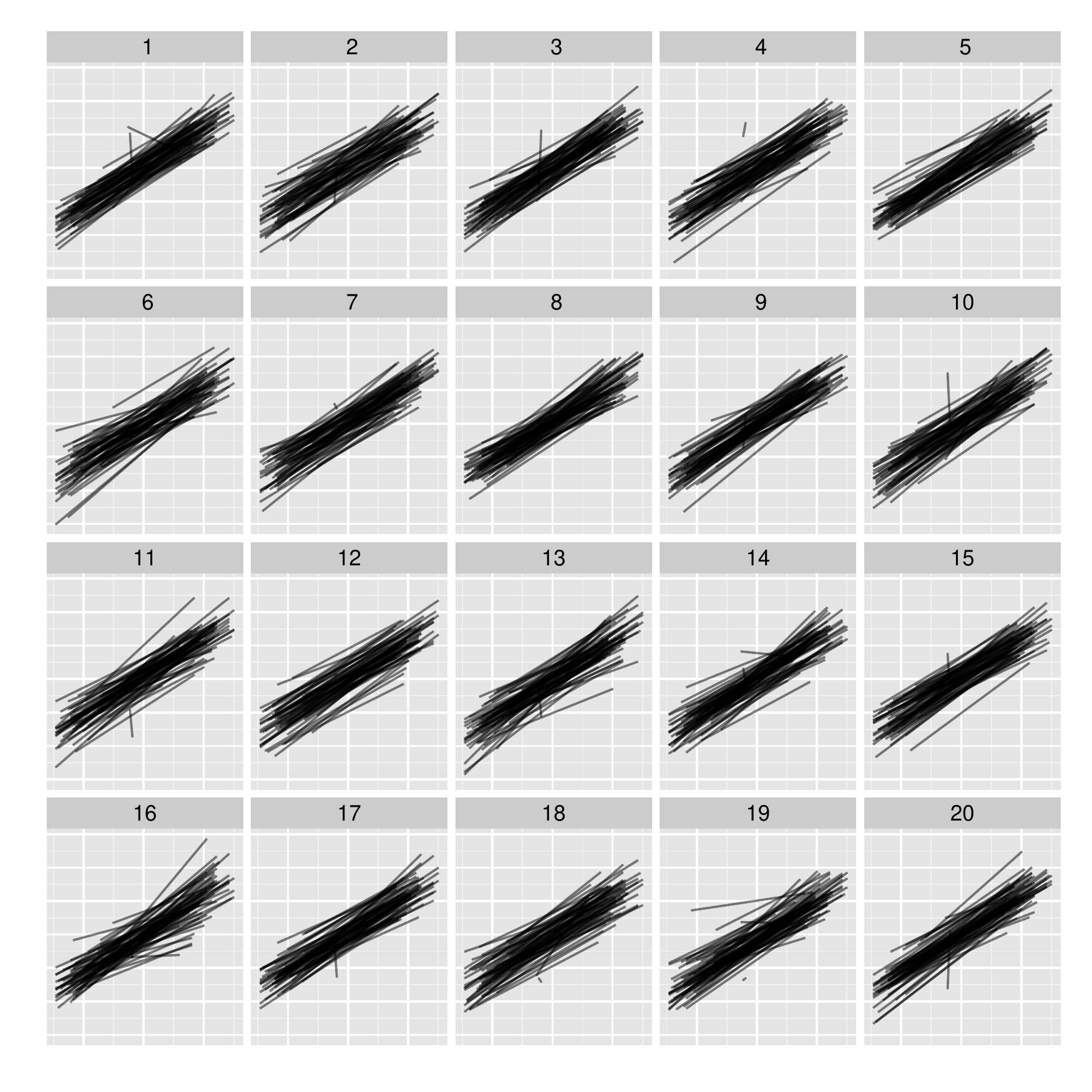}
	\caption{\label{fig:fanned} One of the lineups used in an Amazon Turk study to test for the necessity of random slopes in an LME model. Participants were asked  to identify the most `different' plot and give a reason for their choice. Of the 68 observers, 10 identified the data plot as the most different. This indicates the data are recognizably different from the null plots, indicating the need to include a random slopes component in the model. Further details on data, model and the location of the data plot are given in Section~\ref{sec:random}.}
\end{figure}

Figure~\ref{fig:fanned} shows an example lineup. Each panel presents  line segments of different lengths  with varying slopes. Observers were asked the question `Which plot is the most different?'    Any  information revealing the context of the data, such as axis labels, units, titles, and legends,  was carefully removed to  avoid subjective bias \citep{meilgaard}, ensuring that observers made decisions based purely on the data display. 

Based on a human subject study (described in more detail in Section~\ref{study} of the supplement) run through the Amazon MTurk service \citep{amazon}, 11 out of 73 observers  chose the  data plot, shown in panel  \#($\sqrt{144} + 4$)\footnote{We encode the panel number as a mathematical expression to pose a cognitive obstacle, allowing the reader to evaluate the lineup before being biased by knowing the answer.}
 of the lineup in Figure~\ref{fig:fanned}, resulting in a visual $p$-value of 0.0171. This leads us to reject the null hypothesis that the data plot 
is consistent with the model that generated the data, on which all of the null plots are based. The  model and its corresponding null hypothesis  are explained in detail in Section~\ref{sec:select}.

Unlike classical hypothesis tests, visual inference  allows us to collect additional information on what aspect of the display led each observer to their choice. 
This  information makes it possible to assess which part of the null hypothesis is violated, something not feasible in classical hypothesis tests. For example,  `Spread' as the reason for an observer's choice (over `Outlier', `Trend',  `Asymmetry', or `Other') in Figure~\ref{fig:fanned} was associated with the highest probability of picking the data plot over a null plot (see Table~\ref{tab:reasons} of the supplement).


\section{Issues with LME Models}\label{sec:lme}

In this paper we consider the two-level continuous-response linear mixed-effects model (LME) with uncorrelated errors fit either by maximum likelihood or restricted maximum likelihood. More specifically,
\begin{equation}\label{eq:hlm}
	\underset{(n_i \times 1)}{\bm{y}_i} = \underset{(n_i \times p)}{\bm{X}_i} \ \underset{(p \times 1)}{\bm{\beta}} + \underset{(n_i \times q)}{\bm{Z}_i} \ \underset{(q \times 1)}{\bm{b}_i} + \underset{(n_i \times 1)}{\bm{\varepsilon}_i}
\end{equation}
where there are $i=1,\ldots,g$ non-overlapping groups, $\bm{X}_i$ and $\bm{Z}_i$ are design matrices for the fixed and random effects, respectively, and $\bm{\beta}$ denotes the fixed effects. $\bm{b}_i$ denotes the random effects for group $i$, which are assumed to follow a multivariate normal distribution and are independent between groups. $\bm{\varepsilon}_i$ denotes the error terms, which are assumed to be i.i.d. normal random variables. 


Similar to simple linear models, residuals form the diagnostic core of a LME model. But, LME model residual analysis is complicated by the fact that there are numerous quantities that can be defined as \emph{residuals}, with each residual quantity being associated with different aspects of the model. The two fundamental residuals for model checking considered here are: (1) the error terms (i.e.~level-1 residuals), and (2) the predicted random effects (i.e.~level-2 residuals).

The use of random effects models comes at the cost of complicating model exploration and validation.
Problems addressed in this paper include:

\begin{enumerate}
\item {\em Model selection:}
\begin{enumerate}
\item {\em Assessment of asymptotic distributions}: Test statistics used for model selection and validation rely on asymptotic reference distributions which often perform poorly in finite sample situations. 
For example, the unconditional empirical distribution of the predicted random effects  does not resemble the theoretical distribution unless strong assumptions are met \citep[Theorem 3.2 and Lemma 3.1]{Jiang:1998vt}. In many finite sample situations these assumptions do not hold, making conventional use of Q-Q plots and tests of the empirical distribution function ineffective for distributional assessment \cite[see the supplement of][for supporting simulation results]{adam}.

\item {\em Boundary issues  in the comparison of nested models: } 
When evaluating the significance of terms in the random effects structure, the likelihood ratio test statistic does not have the usual $\chi^2$ reference distribution if we are testing whether a variance component lies on the boundary of the parameter space. This results in tests for the random effects that tend to be conservative.

\end{enumerate}


\item {\em Model checking:} As seen in Figure~\ref{fig:motivation}, residual plots might display noticeable patterns that are artifacts of both the data structure and the model estimation procedure rather than indications of lack of fit. 
This problem is especially pronounced for plots of the error terms, as they often exhibit patterns that appear to be indicative of heteroscedasticity, but are merely consequences of data imbalances or sparsity.

\end{enumerate}

The above issues are well-known, and in special circumstances adjustments to the methodology have been proposed. For example, \cite{Stram:1994wd} suggest using a 50:50 mixture of $\chi^2_q$ and $\chi^2_{q+1}$ when testing $q$ versus $q + 1$ random effects; however, this adjustment is not successful in all cases.  \cite{Lange:1989uu} suggest using weighted Q-Q plots to assess the distributional assumptions made on the random effects. This approach is effective in settings where the residual variance, $\sigma^2$, is small relative to the variance components for the random effects, but breaks down when this is not the case \citep{adam}.
 
In the following sections we expand on the problems encountered in model selection and model checking, respectively, and present solutions utilizing visual inference.

\section{Model selection}\label{sec:select}

\noindent


 Model selection for linear mixed-effects models relies on the comparison of nested models for the selection of both the fixed and random components. 
It is standard practice to use a $t$-test, $F$-test, or likelihood ratio test to determine whether a fixed effect describes a significant portion of the unexplained variability. Alternatively, likelihood-based criteria, such as AIC or BIC, are often used to overcome the problem of being able to only deal with nested models.  
When selecting random effects, likelihood ratio tests are most commonly used. 
However, situations often arise that complicate such tests. 
Some of these situations are outlined below:

\begin{description}
\item[\bf Fixed effects: ] Likelihood ratio tests based on REML estimation cannot be used to test different fixed effects structures. Maximum likelihood estimation allows for such comparisons, but is anti-conservative. 
Defining the appropriate degrees of freedom for  $t$- or $F$-tests provides another  complication in testing scenarios of LME models. Various approximate $F$-tests propose solutions for estimating the degrees of freedom for these tests, but these typically lead to different results  \citep{Verbeke:2000fh}. Inflated Type I error rates and low power  become a problem  in the  approximate $F$-test when there are a few groups \citep{Catellier:2000vr}. \cite{Kenward:1997ft} propose the use of a scaled Wald statistic  that has a sampling distribution that is well approximated by an $F$ distribution; however, this approach has inflated type I error rates in some small sample cases  \citep{Gomez:2005dw}. \cite{Skene:2010kf} propose an alternative to the Kenward-Roger approximation that achieves nominal type I error rates, but this procedure suffers from low power. 
%
%
%

%
\item[\bf Random components: ] When testing for the inclusion of a variance component, the parameter being tested lies on the boundary of the parameter space, and the asymptotic distribution of the likelihood ratio is no longer $\chi^2$. 
Various approaches to deal with this problem have been suggested in the literature. 
\citet[Section 3.5]{Demidenko2013} presents an exact $F$-test of the  null hypothesis $\mathbf{D} = \mathbf{0}$. Unfortunately, the all-or-nothing inclusion of random effects limits the use of this test in practice.
Approximations have been suggested and shown to be useful in many situations \citep{Stram:1994wd, Morrell:1998ua}, but no single approximation holds for all situations. 
Alternatively, the rule of thumb suggested by \citeauthor{Stram:1994wd} can be utilized with the knowledge that the results may be sub-optimal. 
This leads to a need for simulation studies to determine the proper adjustment to the reference distribution in every situation. 
\end{description}

Visual inference provides an alternative to conventional hypothesis tests that does not require different rules based on the method of estimation or location of a parameter in the parameter space, avoids the tricky business of defining degrees of freedom, and allows for the testing of subsets of the random effects. Rather, visual inference depends on the choice of an appropriate plot highlighting the aspect of the model in question, the number of null plots, and the number of independent observers. 

\subsection{Fixed effects} 
To test the significance of a fixed effect, we suggest using a plot comparing a residual quantity from the model without the variable of interest with the values of that variable. 
The residual used depends on the level at which the variable of interest enters the model: if the variable enters at the observation-level (level-1), then the level-1 residuals are used; if the variable enters at the group-level, then both the level-1 and level-2 residuals are explored as the variable has the potential to explain additional variation at either level of the model.
Additionally, the type of plot depends on the variable type---if a continuous variable is targeted,  a scatterplot with a smoother is suitable for testing; for a discrete covariate, we make use of side-by-side box plots. 
In this setting, the null plots are generated using the parametric bootstrap  with a model that omits the variable of interest, details of which can be found in Section~\ref{app:nullplots}. The true plot is constructed from the same model, but is fit to the observed data. 

\begin{figure}[h]
	\centering
	\includegraphics[width=0.8\textwidth]{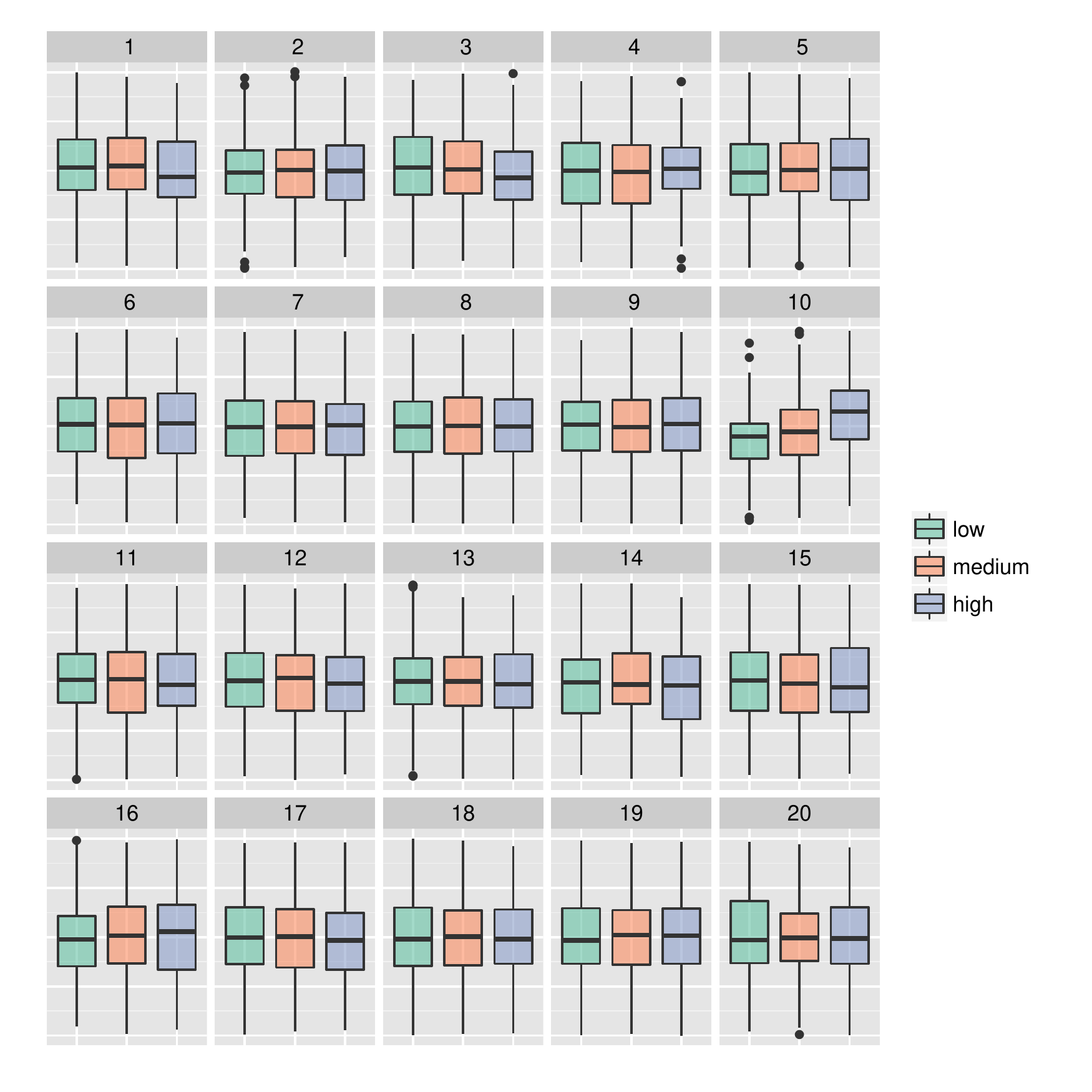}
	\caption{\label{fig:boxplot-ordered} {Lineup of level-1 residuals in side-by-side box plots to test significance of a discrete covariate. Of the 68 participants, 60  identified the data plot from the lineup.  This indicates that the covariate is necessary for the model.}}
\end{figure}

Figure~\ref{fig:boxplot-ordered} illustrates the use of this type of lineup. In our study, 68 observers were asked to choose the plot that is the most different from the rest. Sixty  observers identified the data plot in panel \#($2^3 + 2$), 
with over 90\% pointing to the trend as the distinguishing feature. 
This lineup is chosen
to determine whether a child's language development (low, medium, or high) at age two (plotted on the $x$-axis)  is associated with the development of social skills for children diagnosed with autism spectrum disorder. 
Displayed on the $y$-axis are \mbox{level-1} residuals from a longitudinal model, i.e., the grouping in model~\ref{eq:hlm} is given by each individual. Clearly, language development at age two accounts for a significant amount of the remaining residual variability.


\subsection{Random effects}\label{sec:random}
Tests of the random part of an LME model focus on two questions: (1) whether a marginal random effect improves the model and (2) whether allowing the random effects to be correlated improves the model. Different plots must be used to answer each question. To answer the first question, we suggest using plots comparing the response and the explanatory variable of interest using appropriate (often linear) smoothers for each group. Scatterplots comparing the predicted random effects can be used to answer the second question.

The lineup in Figure~\ref{fig:fanned} was chosen to test the relationship between scores from the General Certificate of Secondary Education Exam (GCSEE) and the  standardized London Reading Test (LRT) (see Section~\ref{data:GCSE}).  Each line segment represents one of 65 inner-London schools. The slope of each line is determined by a linear regression relating the two test scores for each student at a school. 
The question of interest is whether random slopes for LRT scores are required to  represent the relationship between GCSEE and LRT scores ($H_1$). Correspondingly, data for the null plots  were  created by simulating GCSEE scores from a model with the random intercept as its only random effect.
The resulting scores for each school are regressed on LRT scores and model fits are shown as lines.
 If the model is appropriate, then    the observed data should resemble the overall pattern of the lines in the null plots. In this example, we find that the true plot in panel \#($\sqrt{144} + 4$) is identifiable:  11 of the 73 observers picked the data plot, resulting in a visual $p$-value of 0.0171. The main comments participants gave to explain their choice were the spread and trend of the line segments in the plot. This is consistent with a larger variance in the  slopes than the null model allows;
  thus, we find evidence supporting the inclusion of a random slope for standardized LRT. This conclusion agrees with the results of the likelihood ratio test (which shows significance at a level of less than 0.0001), and did not require the use of an asymptotic distribution to calculate the $p$-value. 
  
\begin{figure}
	\centering
	\includegraphics[width = 0.8\textwidth]{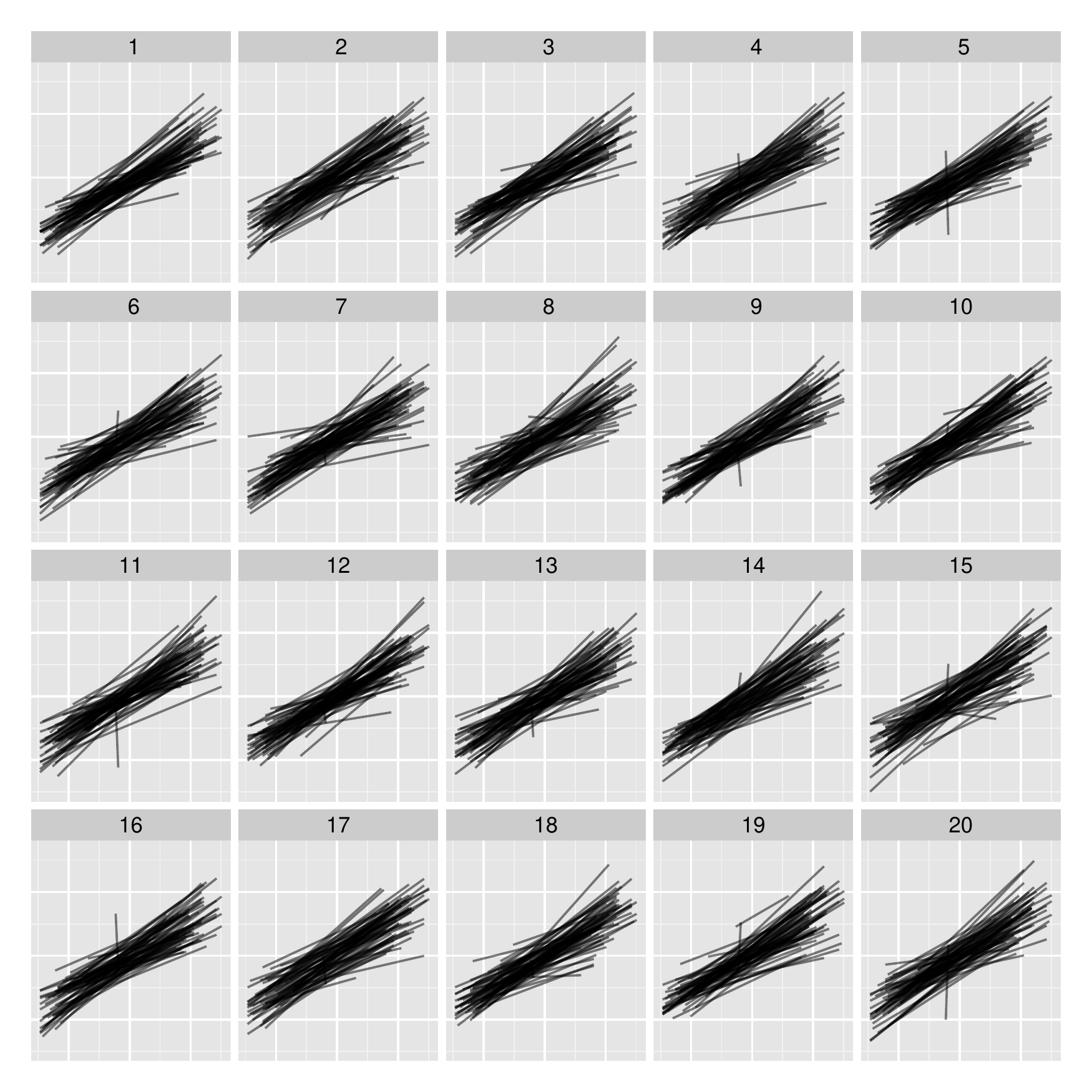}
	\caption{\label{fig:fanned2} A follow-up to Figure~\ref{fig:fanned} where the random slope has been included in the model from which the null plots are created. None of the 64 observers identified the data plot as the most different, providing no evidence of model misspecification. }
\end{figure}

%

{Note that participants are unable to identify the true data from a follow-up lineup (Figure~\ref{fig:fanned2}) in which the random slope was included in the model from which the samples shown in the null plots were created (using a parametric bootstrap as outlined in Section~\ref{app:nullplots}.)
None of the 64 observers identified the data plot in panel \#$(2\cdot 3^2)$, providing no evidence that the covariance structure for the LRT scores is misspecified.


Participants should only see one of Figures~\ref{fig:fanned} or~\ref{fig:fanned2} because both show the same data in the same design. After viewing the first of these figures we cannot, strictly speaking, assume that a participant is still an unbiased judge, because, theoretically, the data panel from the  second lineup could be identified by recognizing it as the same panel that was previously shown. While the chance of this is slim, we only exposed participants to one of a set of dependent lineups.

Having considered the value of a random slope in the model, we next consider whether the model needs to allow correlation between the random effects ($H_1$). While this is an example of a standard likelihood ratio test problem---a correlation of zero is not on the boundary of the parameter space---using a lineup keeps all tests of the random effects in a unified framework. The lineup in Figure~\ref{fig:ranef-corr} shows scatterplots of the predicted random effects with overlaid regression lines. 
 The null plots in the lineup are created by simulation from the model that does not allow for correlation between the random effects, and the true plot is created using the predicted random effects from such a model fit to the observed data.
The slopes of the regression lines are indicative of the amount of correlation. 
 If the correlation between the random effects is not necessary, then the true plot will display little correlation and be indistinguishable from the null plots.
The lineup allows us to gauge the amount of correlation between the random effects while accounting for the effect of shrinkage in the model, avoiding the over-interpretation of structure in such plots discussed by \cite{Morrell:2000ve}.
  In Figure~\ref{fig:ranef-corr}, the true plot in panel~\#($10 + \sqrt{25}$) was identified by 36 of the 63 observers, providing very strong evidence in  support of  the additional parameter for correlation between random effects, which agrees with a $p$-value of 0.0041 from the classical likelihood ratio test.

\begin{figure}[hbt]
	\centering
	\includegraphics[width=0.8\textwidth]{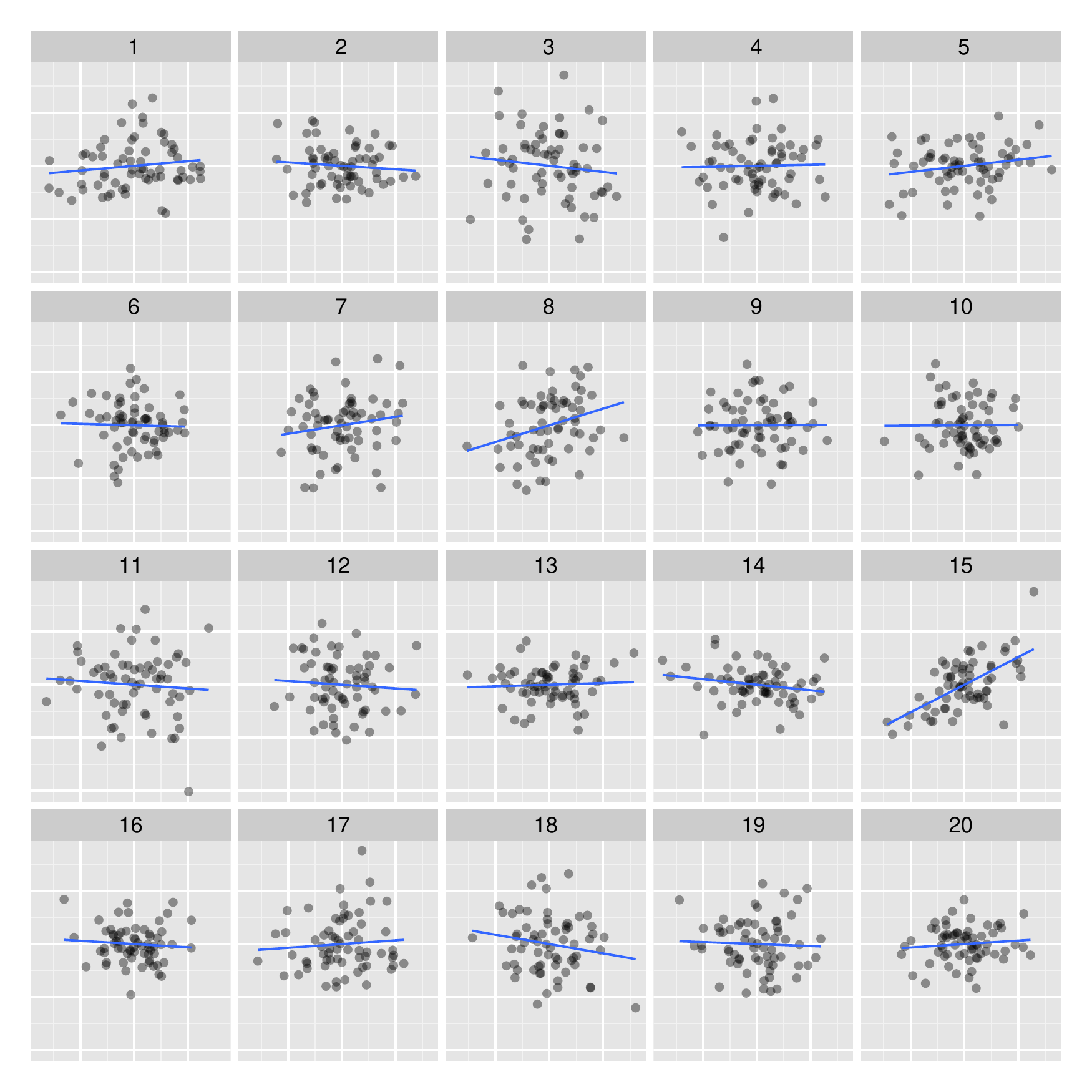}
	\caption{\label{fig:ranef-corr} Lineup for testing the correlation between random effects. Predicted random intercept values are plotted on $x$ against predictions of the corresponding random slopes on $y$. Of the 63 observers, 36 identified the data plot from the lineup, lending strong support for a non-zero correlation between the random effects.  
	}
\end{figure}

\section{Model checking}\label{sec:checking}

In the formulation of model \eqref{eq:hlm} we make a number of assumptions that must be satisfied. In this section we discuss how residual plots can be used with lineups to check the assumptions of homogeneous residual variance, linearity, and normality of the random effects. While we only focus on these assumptions, the discussion is general enough to reveal how visual inference can be extended to check other aspects of the model.

%

\subsection{Homogeneity of residual variance}\label{sec:homogeneity}

Model~\eqref{eq:hlm} assumes homogeneity of the within-group variance. To check this assumption we must verify (a) the homogeneity of the within-group residual variance across the levels of all explanatory variables and (b) check that the within-group variance is also constant between groups. Such investigations are often carried out using plots of the level-1 residuals. In order to guard against mis- or over-interpretation of the residual plots, we, again, employ lineups.
\hfill\\[5pt]
\noindent
\textbf{(a) Checking homogeneity of level-1 residual variance across covariates:}
Residual plots are one of the most-often used tools for checking the relationship between residuals and one of the model's covariates. 
In Figure~\ref{fig:constvar2} we employ a lineup of scatterplots plotting  the level-1 residuals against pressure in the dialyzer study  (see Section~\ref{data:dialyzer}). For the null plots we derive residuals from model-refits of properly specified parametric bootstraps of the investigated model. The plot of the observed data is inserted in position \#($2^4 +3$). Out of 80 evaluations by independent observers the data plot was identified 26 times, providing evidence of heteroscedasticity. 
\begin{figure}[hbt]
	\centering
	\includegraphics[width=0.75\textwidth]{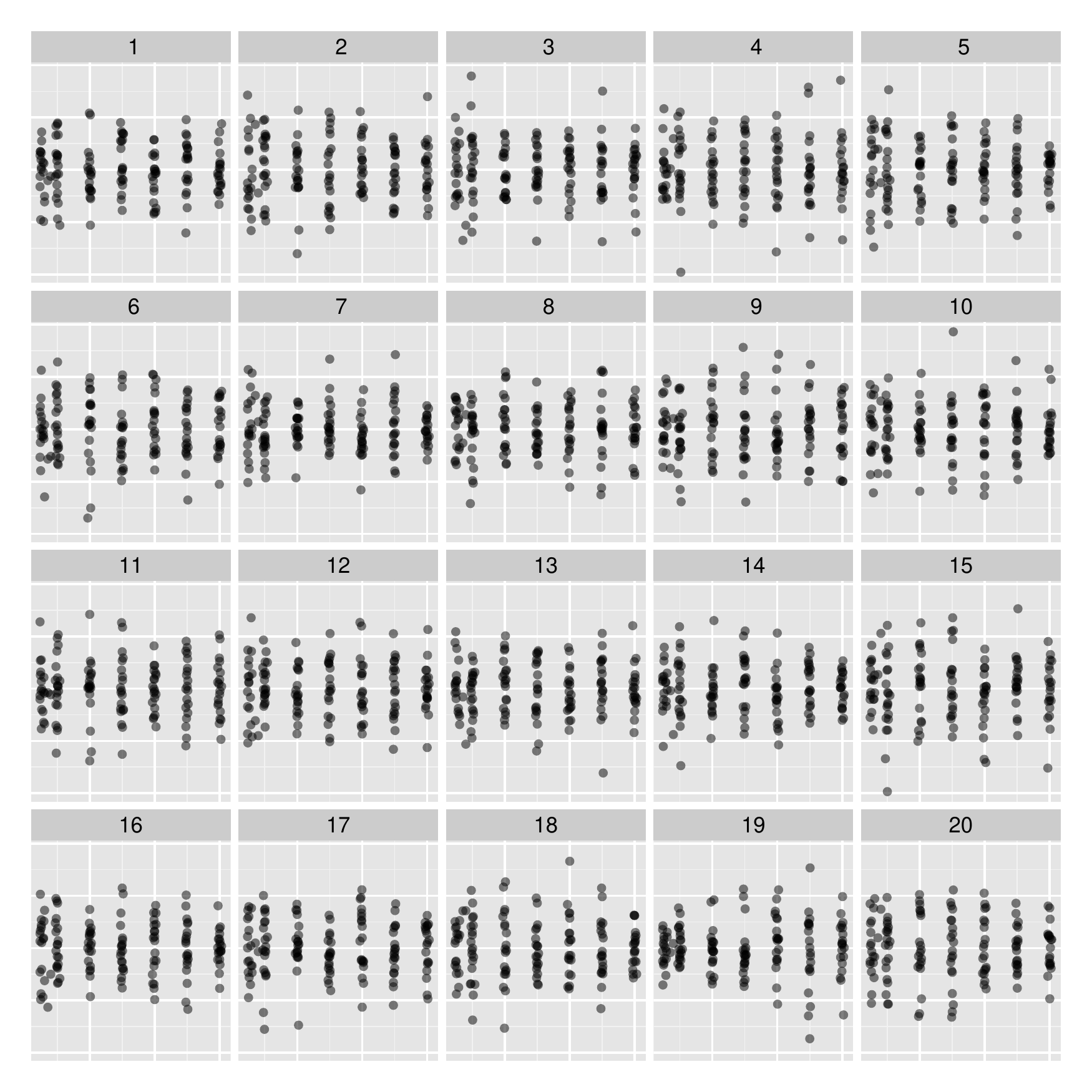}
	\caption{\label{fig:constvar2} 
	Lineup testing homogeneity of the level-1 residuals with respect to  pressure. 
The data plot is in panel \#$2^4+3$. Inspecting only the data plot, we see that an increase in pressure coincides with an increase in the variability of level-1 residuals. The surrounding null plots provide a reference from a properly specified model and thus allow us to assess differences to the data plot.  
	}
\end{figure}
%
%
%
\hfill\\[5pt]
\noindent
\textbf{(b) Checking homogeneity of level-1 residual variance between groups:}
If a covariate is discrete, we  have the choice between different ways of visualizing residuals. When assessing the homogeneity of level-1 residuals across groups, we are, by default, in the situation of comparing continuous values across a discrete range. 
  For the example of  the dialyzer study, we employed side-by-side box plots and side-by-side dotplots to visualize  level-1 residuals against the grouping variable (subjects). 
Figures~\ref{homogeneous-1} and~\ref{homogeneous-2}  in Section~\ref{app:morelineups} show the resulting lineups.
The results from the study clearly support that the design matters, and identify the side-by-side box plot as the far more powerful design: more than one third of all participants (23/61)  identified the data plot from the lineup of box plots, while only one out of 71 participants chose the data plot from the lineup of dot plots. In particular, more than half of the participants noted the spread as their reason for choosing the data plot from the lineup of box plots. 

 Very different group sizes make an assessment of homogeneity in the residual variance more complicated, because this imbalance introduces structures in the plot that are largely artificial. 
%
To overcome this difficulty, we create side-by-side box plots and order groups according to their interquartile range (IQR). Because of their shape,  we have come to call these plots  \emph{cyclone plots}. Figure~\ref{fig:badcyclone} shows a lineup of cyclone plots for 66 patients in a longitudinal study investigating the potential of methylprednisolone to treat patients with severe alcoholic hepatitis (see Section~\ref{data:ahd} for details). The true plot in panel \#($2^3+5$) is easily identified from the field of null plots (by 49 of 73 observers) revealing heteroscedasticity across groups that might not be apparent  in other residual plots.

While the data plot is easily identified, any panel from this lineup considered separately exhibits a structure that might, taken by itself, lead an analyst to the conclusion that the within-group variance varies across the vertical axis. However, placing the true plot into the lineup forces the analyst to consider most of this structure as inherent to the data structure rather than evidence against the hypothesis of homogeneity. The fact that observers are able to identify the data plot indicates that the data plot has additional structure inconsistent with homogeneity. The use of lineups incorporates the comparison of the data to what is expected under a properly specified model, eliminating the subjective interpretations encountered with the use of single plots. 


%
%
%

\begin{figure}[hbt]
	\centering
	\includegraphics[width=0.8\textwidth]{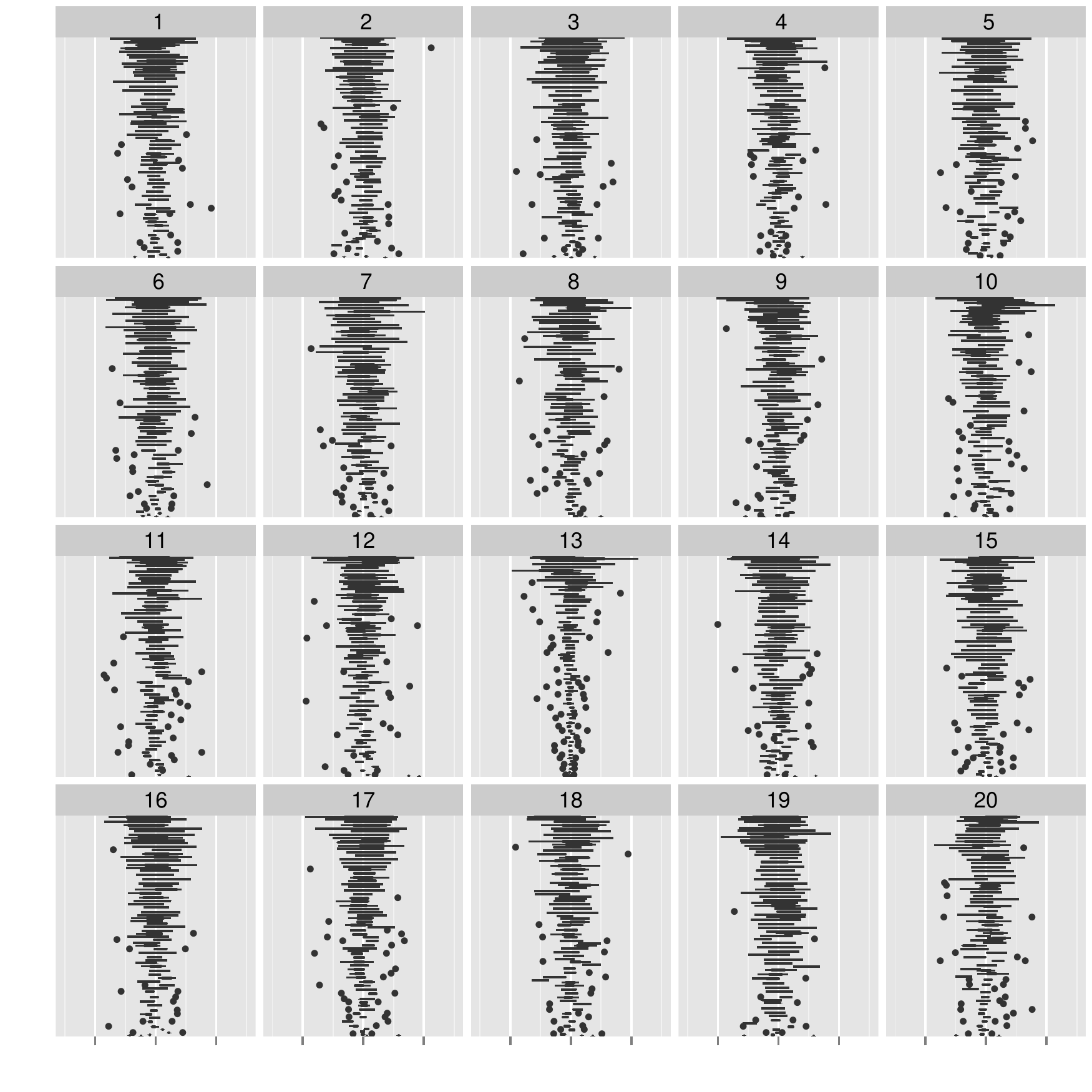}
	\caption{\label{fig:badcyclone} 
	Lineup of cyclone plots to assess homogeneity of the level-1 residuals between groups. Residual values are plotted in horizontal side-by-side box plots by groups. The order of the box plots is determined by the interquartile range of the residuals. 
 Of the 73 observers, 49 chose the data plot as the most different, indicating a significant deviation from the assumptions of the residual distribution in a model of the observed data.} 
\end{figure}

An alternative approach to detect heteroscedasticity of the level-1 residuals across groups is to use a test based on the standardized measure of dispersion given by
\begin{equation}\label{eq:d}
	d_i = \frac{\log\left( s_i^2 \right) - \left[ \sum_i (n_i - r_i) \log\left( s_i^2 \right) / \sum_i  (n_i - r_i) \right]}{\left(2 / (n_i - r_i)\right)^{1/2}},
\end{equation}
where $s_i^2$ is the residual variance within each group based on separate ordinary least squares regressions and $r_i$ is the rank of the corresponding model matrix  \citep{Raudenbush:2002}. The test statistic is then
\begin{equation}
	H = \sum_{i=1}^{g^*} d_i^2
\end{equation}
which has an approximate $\chi^2_{g^*-1}$ reference distribution when the data are normal and the group sizes are ``large enough.'' Here we use $g^*$ because ``small'' groups may be excluded from the calculation as they provide less reliable information about the residual variance (assuming that there are enough observations to fit the model), but this is a subjective choice. A common rule of thumb is to exclude groups with samples sizes smaller than 10. If the distributional assumptions are violated, or we do not have large enough group sizes,  the approximation to the $\chi^2$ distribution breaks down. In the methylprednisolone study each subject was observed at most five times, with 19 subjects dropping out of the study early. Due to the small group sizes the $\chi^2$ approximation is inappropriate, forcing the analyst to rely on simulation to construct the sampling distribution of the test statistic, which is not only computationally more demanding than the generation of 19 null plots, but lacks power with small sample sizes. Additionally, we have found the $H$-statistic to be sensitive to the choice of the minimum group size.
Table~\ref{tab:pvalues} shows the huge difference in conventional $p$-values to those obtained through a bootstrap test (based on 10,000 bootstrap $H$-statistics) for the radon data set.
There is a clear discrepancy between the two $p$-values, indicating a need for simulation-based methods. Additionally, the sensitivity of the simulation-based test to the minimum group size is apparent.  

\begin{table}[ht]
\caption{\label{tab:pvalues} Are error terms homoscedastic? Raudenbush and Bryk's conventional test is highly sensitive to the choice of minimal group size.  
The naive $p$-values are obtained using the $\chi^2$ approximation. Bootstrap $p$-values were calculated from the empirical distribution of the test statistic based on 10,000 simulated statistics obtained using the parametric bootstrap for each minimum group size. There is a clear discrepancy between the two $p$-values indicating a need for simulation-based methods.
} 
\centering
\begin{tabular}{rrrrrrr}
  \hline
 Minimum group size & $H$ & d.f. & Naive $p$-value  & Bootstrap $p$-value \\ 
  \hline
3 & 116.6 & 73 & 0.0009 & 0.9178 \\ 
  4 & 96.8 & 62 & 0.0031 & 0.7980 \\ 
  5 & 77.9 & 45 & 0.0017 & 0.6066 \\ 
  6 & 75.8 & 38 & 0.0003 & 0.5313 \\ 
  7 & 59.0 & 33 & 0.0036 & 0.4697 \\ 
  8 & 51.2 & 29 & 0.0066 & 0.4119 \\ 
  9 & 39.6 & 26 & 0.0426 & 0.3509 \\ 
  10 & 27.7 & 21 & 0.1490 & 0.2595 \\ 
  11 & 26.6 & 19 & 0.1145 & 0.2260 \\ 
  12 & 23.7 & 17 & 0.1281 & 0.1952 \\ 
  13 & 23.7 & 16 & 0.0966 & 0.1873 \\ 
  14 & 8.2 & 11 & 0.6940 & 0.1360 \\ 
  15 & 5.1 & 7 & 0.6429 & 0.0764 \\ 
   \hline
\end{tabular}
\end{table}
If counties with fewer than 10 observations are excluded, the conventional test yields a $p$-value of 0.149; however, if only counties with fewer than 5 observations are excluded, the conventional test indicates strong evidence of heterogeneity based on a $p$-value of 0.0017.
This sensitivity to the group size is a clear weakness of the conventional test, and casts doubt on its usefulness in situations with small group sizes.  
Performing the simulation-based version of the test in this example results in a $p$-value of 0.6066, providing no evidence of heterogeneity. 

  In contrast, consider Figure~\ref{fig:goodcyclone}, another lineup of cyclone plots.  The data underlying this example are radon measurements across counties in Minnesota (see Section~\ref{data:radon}).
Here, level-1 residuals by county are plotted from a model that only includes counties with at least five observations. 
Only one out of  59 participants identified the data plot shown in panel \#$(4^2 - 6)$ from the lineup, providing no evidence against homogeneity. 
The visual test is not overly sensitive to group sizes. Counties with fewer than 5 observations were eliminated because box plots are not appropriate for such small group sizes, but could be included in the representation as dot plots. While we are still slightly constrained by group size, we are far less constrained than with the conventional test, and have a clear way to choose the minimum group size based on our ability to render box plots.

\begin{figure}[hbt]
	\centering
	\includegraphics[width=0.8\textwidth]{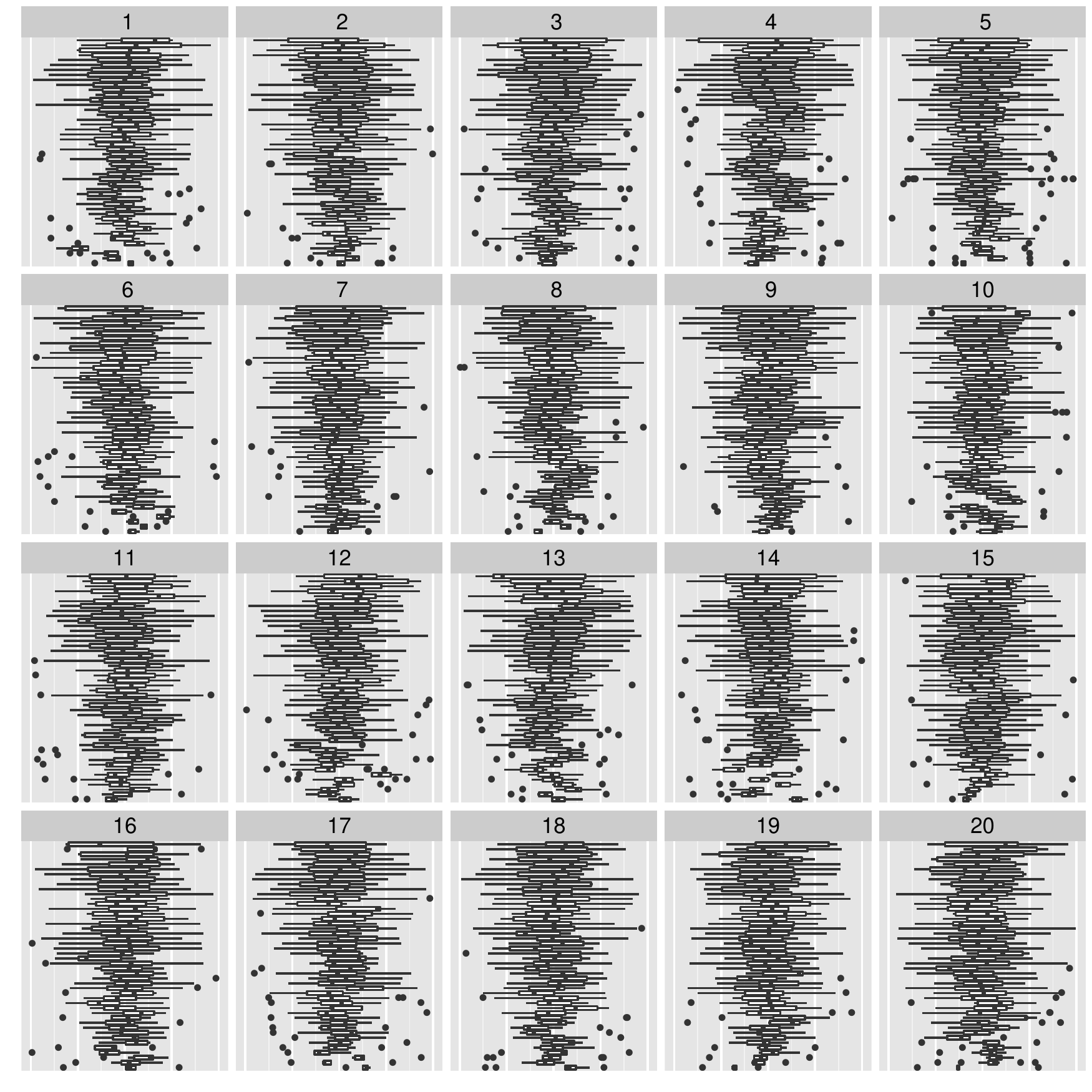}
	\caption{\label{fig:goodcyclone} Lineup of 20 cyclone plots of level-1 residuals used to test the assumption of homogeneous level-1 residual variance.  Only one of 59 observers identified the data plot as the most different, providing no evidence against homogeneity.
	}
\end{figure}

\subsection{Linearity}

Scatterplots with smoothers can also be used to check that the relationship between the explanatory variables and response variable is in fact linear. Figure~\ref{fig:linearity} shows such a lineup testing the linearity of an observation-level explanatory variable. Out of 63 observers, 60 identified the true plot in panel \#($2^3 + 2$), providing evidence that the mean structure is misspecified. This example comes from the dialyzer study and considers a model with only linear and quadratic terms for transmembrane pressure (see Section~\ref{data:dialyzer}). Based on the data panel, it is clear that a higher-order polynomial is required. 
Once a polynomial of degree four is included in the mean structure of the fixed effects, the nonlinear pattern in the residuals is removed. A lineup for the quadratic model is shown in Figure~\ref{fig:constvar2}. In this lineup, the data plot is identified due to heteroscedasticity in the residuals.  This highlights the flexibility of lineups due to the general phrasing of the alternative hypothesis. By tracking observers' reasons for the choice of plot in a lineup, we can distinguish between different alternatives. 
Using the same framework we will get test results based on where we are in the modeling process: as long as the mean structure is not correctly specified, it is most likely the distinguishing feature. Once the mean structure is properly specified, the lineup changes to test for homogeneity of variance. 

\begin{figure}
	\centering
	\includegraphics[width=0.8\textwidth]{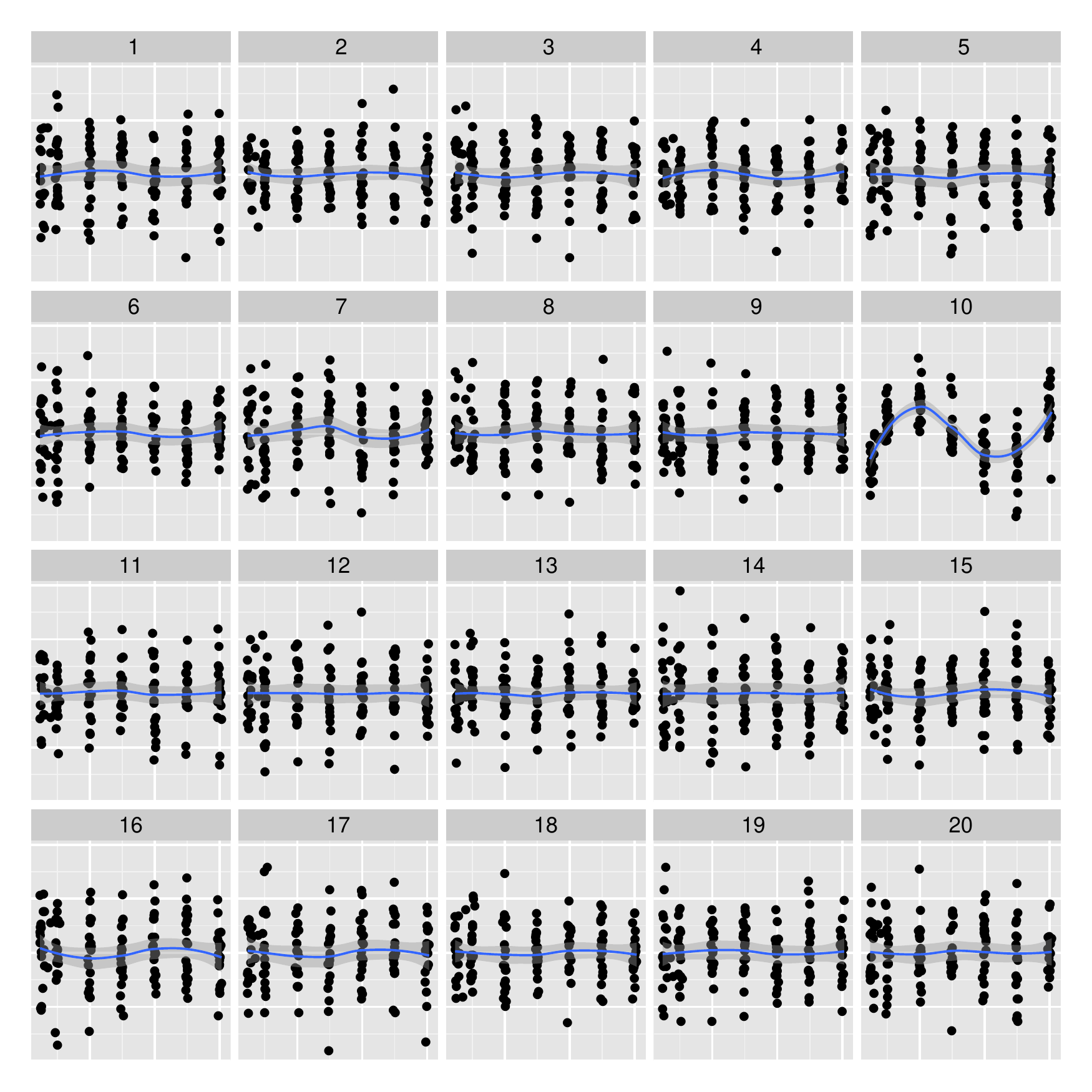}
	\caption{\label{fig:linearity} Lineup testing  for nonlinearity of a covariate. Level-1 residuals (on $y$) from a model of the dialyzer data are plotted against pressure settings (on $x$). Pressure is clearly identified (by 60 out of 62 observers) as an important higher-order covariate of the model.  } 

\end{figure}

To extend checks of linearity to group-level variables we suggest the use of the level-2 residuals. 

\subsection{Distributional assessment}



Recall that in model~\eqref{eq:hlm} we assume that the random effects, $\bm{b}_i$, are a random sample from a multivariate normal distribution 
and are independent from the error terms, $\bm{\varepsilon}_i$, which are assumed to be a random sample from a normal distribution. 
During the model fitting process, however, the predicted random effects are the conditional means of the random effects given the data. While in certain situations the empirical distributions of the residuals in the linear mixed-effects model do converge in probability to their true distributions, very strong assumptions that are not realistic in finite samples are required \citep[Theorem 3.2 and Lemma 3.1]{Jiang:1998vt}. In practice, the predicted random effects (i.e., the level-2 residuals) will not resemble the unconditional distribution; thus, distributional tools such as individual Q-Q plots will lead to erroneous conclusions about the distributional assumptions. Lineups help to overcome this complication.


\begin{figure}[hbt]
	\centering
	\includegraphics[width=0.8\textwidth]{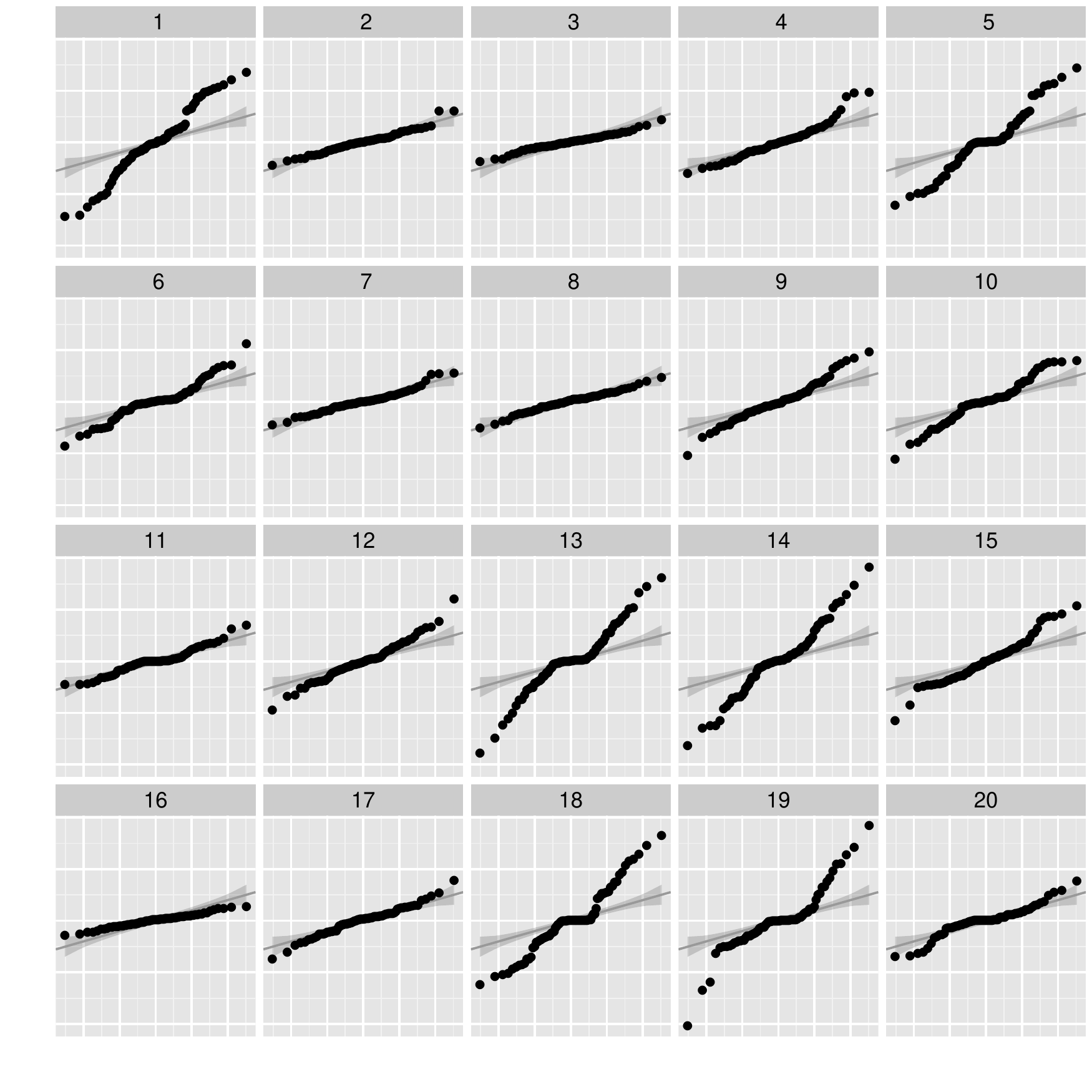}
	\caption{\label{fig:qqlineup-1}
	Lineup of Q-Q plots assessing the distribution of the random slope for the radon data. The lines and confidence bands show the asymptotic distribution. Several of the null plots (and the data plot) exhibit strong deviations from the asymptotic normal distribution. Compared to the null plots, the data plot does not stand out: none of the 65 observers identified the data plot as the most different.
	}
\end{figure}

Figures~\ref{fig:qqlineup-1} and \ref{fig:qqlineup-t} illustrate the use of lineups to test the distributional assumptions in a linear mixed-effects model. 
The null plots for both of these lineups show Q-Q plots of the predicted random slopes  after re-fitting model~\eqref{eq:hlm} to
bootstrap observations generated using the parametric bootstrap as outlined in Section~\ref{app:nullplots}. Consequently, the null plots represent Q-Q plots of the random slopes for a properly specified model.
In Figure~\ref{fig:qqlineup-1}, the Q-Q plot of the predicted random slopes of model~\eqref{eq:hlm} fit to the radon data was inserted into the lineup, while the lineup in Figure~\ref{fig:qqlineup-t} included a Q-Q plot of the random slopes in model~\eqref{eq:hlm}
where the random effects were simulated from a multivariate $t_3$-distribution.
In both lineups, Q-Q plots are drawn with lines representing the asymptotic normal distribution and shaded confidence bands. It is obvious that in many of the panels the empirical distribution of the predicted random effects---both for the null plots and true plot---does not align with the asymptotic distribution.
This is particularly pronounced in this example because the radon data consist of groups of very uneven sizes, resulting in a high degree of shrinkage.
While the confidence bands show the relationship of the predicted random effects to the hypothesized distribution, this is known to be an ill-conceived comparison. 
However, the main comparison in lineups is not a comparison of the predicted random effects to the normal distribution, but rather a comparison of the empirical distribution of the random effects between the null and observed plots. 
Consequently,  conclusions drawn from the lineups relate to evidence of consistency between the true plot and what is expected under a properly specified model. For example, the true plot in panel \#($2^4 - 6$) in Figure~\ref{fig:qqlineup-1}  is indistinguishable from the null plots (none of 65 observers identified this plot), providing no evidence of a violation of (asymptotic) normality; however, when compared only to the normal distribution, the observed Q-Q plot would 
be rejected by any standard test for normality (e.g., the $p$-value of the Anderson-Darling test is .0004 for the data panel).
Panel \#19 was identified  most often from the lineup in Figure~\ref{fig:qqlineup-1}; it was picked by 29 out of 65 observers. Other panels  selected at least four times were \#1, 13, 14, 16, and 18. 
This example also shows the impact of confounding between the different levels of the mixed effects model on the distribution of the predicted random effects \citep{adam}: even for null plots, where the random effects were generated from a normal distribution before re-fitting the model, the predicted random effects do not look normal. In fact, the Anderson-Darling test of normality rejects the null hypothesis of normality for 16 of the null plots at the 0.05 significance level.

\begin{figure}
	\centering
	\includegraphics[width=0.8\textwidth]{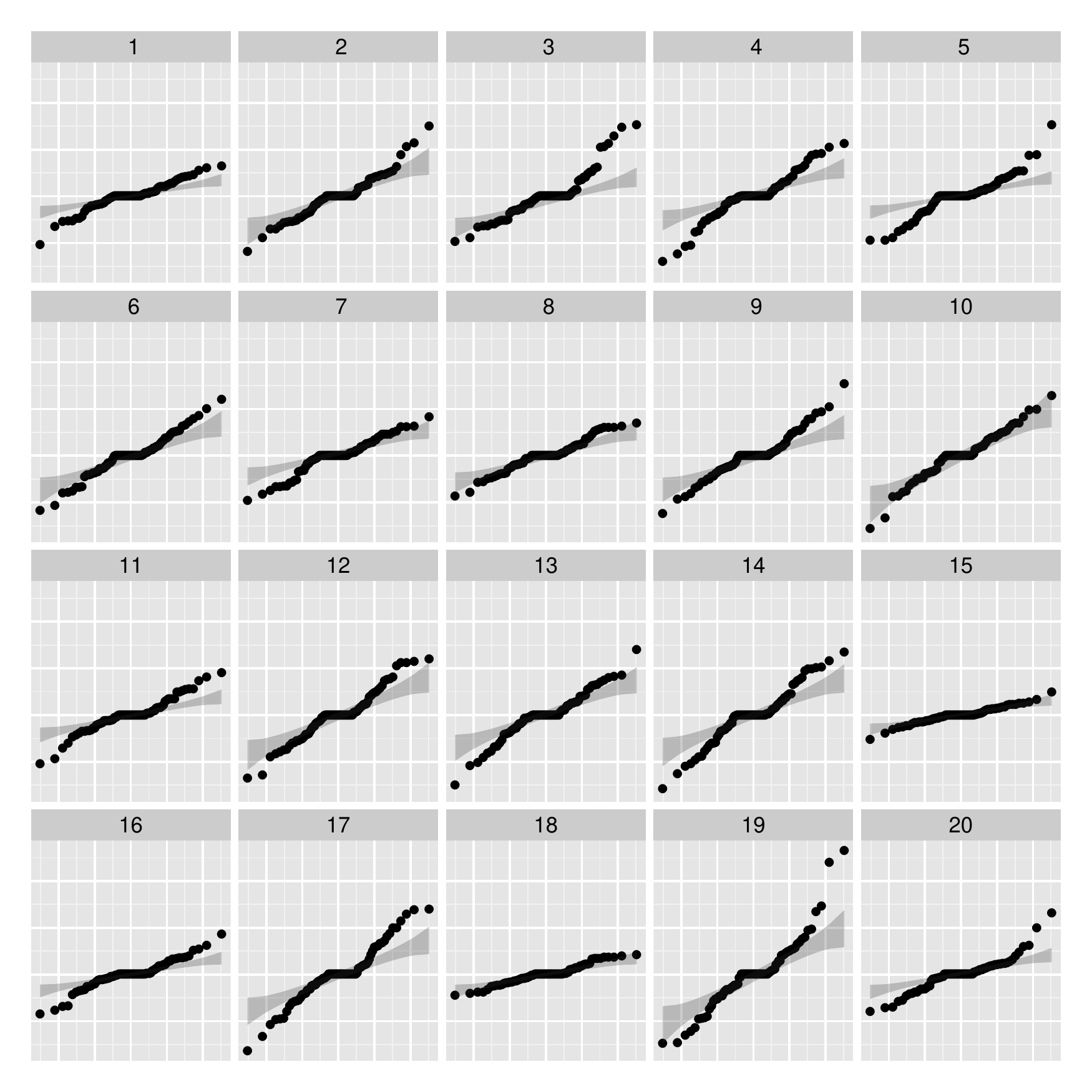}
	\caption{\label{fig:qqlineup-t} Lineup of Q-Q plots assessing the distribution of random slopes. For the data plot,  the random effects are simulated from a multivariate $t_3$ distribution; the null plots show model re-fits of data with normally distributed random effects. Of the 63 observers, 26 identified the data plot as the most different from the lineup, providing evidence that  we can visually detect a violation of normality of the random effects.
	}
\end{figure}

To 
demonstrate that lineups are indeed able to detect distributional violations, we constructed another lineup where the ``true plot'' was based on predicted random effects obtained from a model re-fit to data simulated using normal error terms and $t_3$ random effects.
The null plots display Q-Q plots for the random slopes extracted from this simulated model.
 This lineup is shown in Figure~\ref{fig:qqlineup-t}. Twenty six out of 63 observers identified the true plot in panel \#($\sqrt{49} + 3\cdot4$), providing evidence 
that we can visually detect a violation of normality of the random effects in Figure~\ref{fig:qqlineup-t}. 
This indicates that lineups of Q-Q plots provide an avenue for distributional assessment where conventional methods fail. Further investigation is needed to explore limitations of this approach. 
 The ability of a lineup to distinguish a $t$ distribution for the random effects in the radon study shows that the approach has fewer limitations than conventional approaches, justifying our preference.

\section{Protocol}


Throughout this paper we have illustrated how visual inference can help select and diagnose LME models. In all of the examples presented, we have followed the same basic protocol:

\begin{enumerate}
\item {\bf Create lineup data:} Assuming that the proposed model (for diagnostics) or reduced model (for selection) are correct, we use a parametric bootstrap to simulate new responses, re-fit the model to these simulated responses, and extract the residuals of interest from the proposed model. For each lineup, this process is used to obtain $m-1=19$ simulated null data sets. (See Section~\ref{app:nullplots}.)

\item {\bf Render lineups:}  Draw small multiples of each of the null data sets and randomly insert the observed data among the nulls. Each plot is labeled by a number from 1 to $m$. These IDs are used for identification  and later evaluation of results.

\item {\bf Evaluate lineups:} Present the lineups to independent observers, instructing them to identify the plot most different from the set and asking them what feature led to their choice. These choices came in the form of four suggestions (in checkboxes) and one text box for a free-form answer. 

\item {\bf Evaluate the strength of evidence:} For a lineup of size $m=20$ that has been evaluated by $K$ independent observers, the number of evaluations of a lineup in which the observer identifies the data plot, $Y$, has a Visual distribution $V_{K,m,s=3}$ as defined by \citet{hofmann:2015}.

\end{enumerate}

In practice, the modeler will not give every lineup rendered to a panel of independent observers. Rather, during the model building process many lineups will be rendered with the modeler blinded to the true plot. These lineups replace the traditional exploratory and diagnostic plots traditionally used, providing additional ``protection'' against structure introduced through the model fitting procedure. For critical decisions---perhaps, to finalize model selection or diagnose borderline situations---recruiting a number of independent observers is warranted. While we used the MTurk service \citep{amazon}, modelers could also use their colleagues, provided that they are making these evaluations independently.

For each of the lineup designs described in the paper we constructed five replicates consisting of the same data plot and different sets of nineteen null plots for a total of 75 different lineups. 
487 participants were recruited through the MTurk service \citep{amazon}, each participant was asked to evaluate ten lineups, for a total of altogether 4927 evaluations. 
In addition to the plot choice and rationale, the time taken to answer was recorded and  observers were asked for their confidence level (on a scale from 1=weak to 5=high). Observers were also asked to provide their demographics: age category, gender,  education range, and geographic location (from parts of their ip address). It has been shown that demographics and educational background of observers, while significant, do not have a practical impact on  detection rates of the true plot from a lineup  \citep{majumder:2014}. 

The MTurk service provides access to a large pool of participants, whose demographics reflect those of internet users (i.e.\ relative to the average U.S.\ population, participants are younger and more highly educated; the proportion of female participants is typically around 40\%). To be eligible to participate in the study, we required MTurk workers to have at least 100 completed HITs (Human Intelligence Tasks) with a success rate of at least 90\%.  Assuming that these workers  have a vested interest in their reputation,  we hope that such a selection reduces the rate of participants gaming the system. The pool of MTurk workers is large enough to allow for very fast collection of lineup evaluations; all of the data presented here were collected within a few hours.

While we do not require any statistical training from participants, training in mathematical reasoning has been found to significantly help with performance on lineup tests \cite[see][]{susan:infovis}.

\section{Discussion}\label{sec:conclusion}

We have presented a graphical approach to model selection and diagnosis using lineups constructed by simulation from the model. 
Lineup tests provide us with the framework to test hypotheses and also allow a  subsequent exploration of the plots in the lineup  for additional insight into the data structure. Thus, instead of simply rejecting the null hypothesis, the use of lineups also allows us to explore \emph{why} we are rejecting the null hypothesis. This approach relies  on the simulation process, the design of the graphics created, and observers, but avoids the reliance on asymptotic reference distributions; thereby circumventing the pitfalls of many commonly used tests. While the value of lineup tests is, perhaps, most obvious in such problematic situations, the approach provides an avenue for testing in all situations, even those in which commonly used tests are appropriate.


The graphical approach is relatively new and involves working with human observers recruited on the web. There is a vast experience in statistics in engaging subjects in a traditional lab setting where researchers are closely engaged with the experiment. With MTurk, researchers are working with subjects in a long distance relationship, and there are some participants who will try to ``game the system.'' Nevertheless, results are promising, as MTurk experiments have replicated studies conducted in the traditional lab setting~\citep{crump2013}. Specifically, \citet{heer:2010} achieved matching results to those of \citet{cleveland:1984} using MTurk, and \citet{kosara:2010} found similar results between a lab study and one performed using MTurk.

As with conducting surveys, avoiding leading questions is very important. For this study, observers were generically asked to pick the plot with the most different features. Contextual information, such as labels, axis tick marks, and titles, were removed to avoid subjective bias. Care was taken in constructing lineup sequences and different plot designs so that observers saw the data only once. Because of the finite nature of possible comparison in a lineup, multiple replications of lineups (e.g., five, as in this study) are recommended using different sets of $m-1$ null plots. 

By asking observers for the distinguishing feature(s) of the plot they chose,  graphical tests also provide us with information about the specific violation(s) of the null that is (are) captured in a general alternative hypothesis. A fixed selection of reasons were provided for observers to choose from, but they could also enter a different reason in a text box.  This allowed us to examine responses by reasons for selection to assess individual sub-hypotheses and investigate different types of errors. For example, in Figure~\ref{fig:qqlineup-t} some participants chose plots because they showed the ``most straight line'' or ``closest fit'' to the line. These reasons would indicate that the person may not have had prior experience in reading Q-Q plots, and mistakenly looked for compliance with a theoretical distribution. This can be interpreted as a Type II error. Type III errors \citep{mosteller:1948}, where the null hypothesis is rejected for the wrong reason, can also be detected by investigating participants' choices.

Plot design is important---some designs are better at revealing anomalies than others. At a population level, the results obtained from the same plot design are very stable. 
In this paper, design choices were made using our best judgment based on  experience and results from cognitive psychology. As more studies are conducted more about the power of designs for model diagnostics will be learned, enabling more informed decisions about best practices \citep{adam:tas, infovis:2012}. A further benefit of using the graphical framework for testing is that graphics adapt relatively well to big data situations \citep{unwin:million}. This provides us with a viable approach to assess the practical relevance of a result versus results that show statistical significance purely based on the dimension of the problem. Barring bad design choices, graphical tests allow us to judge practical relevance of a result as ``If we do not see it, it might be there but not be relevant.'' 

While we have tried to present a variety of different plots throughout this paper, one of the strengths of the graphical approach is that a single visualization often provides insight into different aspects of the model. For example, a plot of a continuous covariate and the  residuals of a corresponding model enables us to investigate the presence of a non-linear trend,  the amount of heteroscedascity  (with respect to that covariate), and, to a degree, reveals features of the marginal distribution of the residuals, such as its skewness. The lineup protocol allows us to simultaneously assess several model assumptions. Very few conventional tests allow for this. The paper by \citet{pena:2006} is a rare exception: here, the authors present a test for the global validation of assumptions in a linear model. Alas, the test does not allow for a ready extension to the hierarchical model.

The diagnostics in this paper draw from various sources. Some of the diagnostics presented here are well-known diagnostic  tools, such as Q-Q plots or  scatterplots of residuals with trend lines as suggested by \citet{Cook:1999} for ordinary least squares regression. Some diagnostics are suggestions from the literature specific to mixed effects models. An example of a new diagnostic addressing a  practical need are the cyclone plots of Figure~\ref{fig:badcyclone}. The overarching purpose of  these examples is to show visual diagnostics in a wide variety of situations that all need special consideration in the conventional hypothesis testing setting, but that all fit within the same graphical inference framework.  Many situations, such as outlier detection, were not discussed in this paper, but the results also extend to these problems. The reader is encouraged to examine more examples provided by \citet{Buja:2009hp} and \citet{mahbub:2013}, as well as \citet{RoyChowdhury:2014} for biological applications.

\section*{Acknowledgments}
\if0\blind
{
We would like to thank Mahbubul Majumder and Eric Hare, who conducted the Amazon MTurk study (\url{mahbub.stat.iastate.edu/feedback_turk11/homepage.html}). R \citep{R} was used to implement all analyses and methods discussed in this paper: lme4 \citep{lme4} and HLMdiag \citep{Loy:JSS} were used for model fitting and  calculation of diagnostics, respectively.  ggplot2 \citep{ggplot2}, nullabor \citep{nullabor}, and gridSVG \citep{gridSVG} provided the basis for the visualizations.
This work was funded in part by National Science Foundation grant DMS 1007697. All studies were conducted with approval from the internal review board IRB 10-347.
} \fi

\if1\blind
{
R \citep{R} was used to implement all analyses and methods discussed in this paper: lme4 \citep{lme4} and HLMdiag \citep{HLMDiag, Loy:JSS} were used for model fitting and  calculation of diagnostics, respectively.  ggplot2 \citep{ggplot2}, nullabor \citep{nullabor}, and gridSVG \citep{gridSVG} provided the basis for the visualizations.
} \fi


\bigskip
\begin{center}
{\large\bf SUPPLEMENTARY MATERIAL}
\end{center}

\begin{description}

\item[supplement.pdf] The supplement materials include a description of the data sets used throughout this paper, the experimental setup, and the calculation of $p$-values. Additional lineups that were omitted for brevity are also included. (PDF)
\item[study.csv] Anonymized  results from the Amazon Turk study. (CSV)
\item[analysis.r] R script reproducing the results. (R)

\end{description}



\bibliographystyle{apalike}
\bibliography{hlmviz_bib}

\clearpage

  \bigskip
  \bigskip
  \bigskip
  \begin{center}
    {\LARGE\bf Model Choice and Diagnostics for Linear Mixed-Effects Models Using Statistics on Street Corners (Supplementary Materials)}
\end{center}
  \medskip
  
  \spacingset{1.45}

\begin{appendix}
The materials in this document supplement the information presented in the manuscript 
``Model Choice and Diagnostics for Linear Mixed-Effects Models Using Statistics on Street Corners.'' Section~\ref{supp:model} gives an overview of the theoretical framework of linear mixed effects models. Section~\ref{supp:datasets} describes the data sets used to illustrate the use of visual inference in the paper.
Section~\ref{study} details the experimental setup, generation of null plots, the calculation of $p$-values, and additional lineups that were included in the study, but omitted from the paper for brevity. The figure numbering carries on from the paper for ease of reporting results.

\section{Model overview}\label{supp:model}

Linear mixed-effects models can account for dependence structures when data are composed of groups.
Such structures occur, for example, when individuals are naturally grouped by organization (e.g., students within schools), geography (e.g., voters within states), or design (e.g., respondents assigned to  interviewers). The models allow data to be incorporated at both the observation-level (level 1) and the group-level (level 2, or higher) while also accommodating dependencies between individuals within the same group. 

For data organized in $g$ groups, consider a continuous response linear mixed-effects model (LME model) for each group $i$, $i=1, \ldots, g$:
\begin{equation}\label{eq:hlm}
	\underset{(n_i \times 1)}{\bm{y}_i} = \underset{(n_i \times p)}{\bm{X}_i} \ \underset{(p \times 1)}{\bm{\beta}} + \underset{(n_i \times q)}{\bm{Z}_i} \ \underset{(q \times 1)}{\bm{b}_i} + \underset{(n_i \times 1)}{\bm{\varepsilon}_i}
\end{equation}
where $\bm{y}_i$ is the vector of outcomes for the $n_i$ level-1 units in group $i$, $\bm{X}_i$ and $\bm{Z}_i$ are design matrices for the fixed and random effects, respectively, $\bm{\beta}$ is a vector of $p$ fixed effects governing the global mean structure, $\bm{b}_i$ is a vector of $q$ random effects describing the between-group covariance structure, and $\bm{\varepsilon}_i$ is a vector of level-1 error terms accounting for the within-group covariance structure. The random effects,  $\bm{b}_i$,  are assumed to be a random sample from $\mathcal{N}(\bm{0},\ \bm{D})$ and independent from the level-1 error terms,  $\bm{\varepsilon}_i$, which are assumed to follow a $\mathcal{N}(\bm{0},\sigma^2 \bm{R}_i)$ distribution. 
Here, $\bm{D}$ is a positive-definite $q \times q$ covariance matrix and $\bm{R}_i$ is a positive-definite $n_i \times n_i$ covariance matrix.
Finally, it is assumed that all between group effects have a covariance of zero.

Inference typically centers around either the marginal or conditional distribution of $\bm{y}_i$, depending on whether global or group-specific questions are of interest.
Based on model \eqref{eq:hlm} the marginal distribution of $\bm{y}_i$ for all $i = 1, \ldots, g$ is given by
\begin{equation}\label{eq:marginalmod}
\bm{y}_i \sim \mathcal{N}\left(\bm{X}_i\bm{\beta},\ \bm{V}_i \right),
\end{equation}
where $\bm{V}_i = \bm{Z}_i \bm{DZ}_i\trans + \sigma^2 \bm{R}_i$, and the conditional distribution of $\bm{y}_i$ given $\bm{b}_i$ is defined as
\begin{equation}\label{eq:conditionalmod}
\bm{y}_i | \bm{b}_i \sim \mathcal{N}\left(\bm{X}_i\bm{\beta} + \bm{Z}_i \bm{b}_i, \ \sigma^2 \bm{R}_i \right).
\end{equation}

Similar to simple linear models, residuals form the diagnostic core of a LME model. But, LME model residual analysis is complicated by the fact that there are numerous quantities that can be defined as \emph{residuals}, with each residual quantity being associated with different aspects of the model. The two fundamental residuals for model checking considered here are:
\begin{itemize}
\item the \emph{level-1 (observation-level) residuals}, ~the conditional residuals or error terms: $\widehat{\bm{\varepsilon}}_i = \bm{y}_i - \bm{X}_i \widehat{\bm{\beta}} - \bm{Z}_i \widehat{\bm{b}}_i$,

\item and the \emph{level-2 (group-level) residuals}, ~the predicted random effects $\widehat{\bm{b}}_i$
\end{itemize}
where $\widehat{\bm{\beta}}$ is an estimate of the fixed effects, 
\begin{equation}\label{eq:glsb}
	\widehat{\bm{\beta}} = 
	\left(\sum^g_{i=1} \bm{X}\trans_i \bm{V}^{-1}_i \bm{X}_i \right)^{-1} 
	\sum^g_{i=1} \bm{X}_i\trans \bm{V}_i\inv \bm{y}_i,
\end{equation}
and $\widehat{\bm{b}}_i$ are predictions of the random effects, given as
\begin{equation}\label{eq:eb}
	\widehat{\bm{b}}_i = \bm{D} \bm{Z}_i\trans \bm{V}_i^{-1} 
	\left(\bm{y}_i - \bm{X}_i \widehat{\bm{\beta}} \right), \ \ \ \ \forall\ i = 1, \ldots, g.
\end{equation}
When $\bm{V}_i$ is unknown, estimates for the covariance matrices are used in the above equations. These estimates are commonly found through maximum likelihood (ML) or restricted maximum likelihood (REML).

\section{Data sets}\label{supp:datasets}

All of the data sets used in this paper are publicly available: the General Certificate of Secondary Education Exam data set is available in the R package mlmRev \citep{mlmRev}; the Dialyzer data set is available in the R package MEMSS \citep{MEMSS}; all other data sets can be found in the R package HLMdiag \citep{HLMDiag}.

\subsection{General certificate of secondary education exam data}\label{data:GCSE}

We make use of a subset of examination results of 4,065 students nested within 65 inner-London schools discussed by \cite{Goldstein:1993wm}. The original analysis explored school effectiveness as defined by students' performance on the General Certificate of Secondary Education Exam (GCSEE) in both mathematics and English. This exam is taken at the end of compulsory education, typically when students are 16 years old.  To adjust for a student's ability when they began secondary education, the students' scores on the standardized London Reading Test (LRT) and verbal reasoning group (bottom 25\%, middle 50\%, or top 25\%) at age 11 were recorded. Additional information contained in the data set includes student gender, school gender, and the average LRT intake score for each school.

\subsection{Autism study}\label{data:autism}
In an effort to better understand changes in verbal and social abilities from childhood to adolescence, \cite{Anderson:2007cl, Anderson:2009in} carried out a prospective longitudinal study following 214 children between the ages of 2 and 13 who had been diagnosed with either autism spectrum disorder or non-spectrum developmental delays at age 2. The Vineland Adaptive Behavior Interview survey was used to assess each child's interpersonal relationships, play time activities, and coping skills, 
from which the Vineland Socialization Age Equivalent (VSAE) was computed as an overall measure of a child's social skills. Additionally, expressive language development at age 2 was assessed using the Sequenced Inventory of Communication Development (SICD) and the children were classified into three groups (high, medium, or low). Assessments were made on the children at ages 2, 3, 5, 9, and 13, however, not all children were assessed at each age. Additional information collected on each child includes: gender, race (white or non-white), and initial diagnosis at age 2 (autism, pervasive development disorder (pdd), or non-spectrum). We restricted attention to models concerned with the changes in social skills for subjects diagnosed with autism spectrum disorder having complete data. This results in a reduced data set of 155 children. For more detailed analyses we refer the reader to \cite{Anderson:2007cl, Anderson:2009in}.

\subsection{Methylprednisolone study}\label{data:ahd}

\cite{Carithers:1989} conducted a four week longitudinal study to investigate the effectiveness of methylprednisolone to treat patients with severe alcoholic hepatitis. The researchers randomly assigned 66 patients to receive either methylprednisolone (35 patients) or a placebo (31 patients). Over the study duration, each subject's serum bilirubin levels (in $\mu$mol/L) were measured each week, with the first measurement taken at the start of the study (week 0).

\subsection{Dialyzer study}\label{data:dialyzer}

\cite{Vonesh:1992us} describe a study characterizing the water transportation characteristics of 20 high flux membrane dialyzers, which were introduced to reduce the time a patient spends on hemodialysis. The 20 dialyzers were studied \emph{in vitro} using bovine blood at flow rates of either 200 or 300 ml/min, and the ultrafiltration rate (ml/hr) for each dialyzer was measured at seven transmembrane pressures (in mmHg). \cite{Vonesh:1992us} use nonlinear mixed-effects models to analyze these data; however, they can be modeled using polynomials in the linear mixed-effects framework \citep[see][Section 9.5]{Littell:2006}.

\subsection{Radon study}\label{data:radon}

The data consist of a stratified random sample of 919 owner-occupied homes in 85 counties in Minnesota. For each home, a radon measurement was recorded (in log $pCi/L$, i.e., log picoCuries per liter) as well as a binary variable indicating whether the measurement was taken in the basement (0) or a higher level (1). Additionally, the average soil uranium content for each county was available. The number of homes within each county varies greatly between counties ranging from one home to 116 homes, with 50\% of counties having measurements from between 3 and 10 homes. \cite{Gelman:2006ue} suggest a simple hierarchical model allowing for a random intercept for each county and a random slope for floor level. This is the model from which we simulate predicted random effects.

\section{Experimental Setup and Results}\label{study}
\subsection{Experimental Setup and Calculation of $p$-values}\label{sec:pvalues}
For each of the lineup designs described in the paper we constructed five replicates consisting of the same data plot and different sets of nineteen null plots for a total of 75 different lineups. These were evaluated by 487 participants in altogether 4927 evaluations. For each lineup, observers were instructed to identify the plot  most different from the set and asked what feature led them to their choice. 
These choices came in the form of four suggestions (in checkboxes) and one text box for a free-form answer. For each lineup the time taken to answer was recorded and  observers were asked for their confidence level (on a scale from 1=weak to 5=high). Observers were also asked to provide their demographics: age category, gender,  education range, and geographic location (from parts of the ip address). 

The results of the evaluations for all lineups  are displayed in Table~\ref{tab:results} and the observers' reasons for identifying plots are summarized in Table~\ref{tab:reasons}. 
The significances in Table~\ref{tab:results} are based on the number of evaluations and the number of times that the data plot was identified. For that, we the introduce the random variable $Y$ as the  number of evaluations of a lineup in which the observer identifies the data plot. Assume that the lineup has size $m=20$, and it is shown to a total of $K$ independent observers.  Then $Y$ has a Visual distribution $V_{K, m, s=3}$ as defined in \citet{hofmann:2015},
where $s$ delineates scenario III---i.e., the same lineup is shown to all $K$ observers. The $p$-values in Table~\ref{tab:results} for each of the five replicates are calculated this way.
For the overall $p$-value, we use a simulation based approach to combine the five results. Treating the number of evaluations $(K_1, ..., K_5)$ as fixed, we simulate assessments of lineups without signal as follows: we assume that the signal in a plot is complementary to its $p$-value, which is i.i.d. $U[0,1]$ under the null hypothesis. We further assume that the probability an observer picks a  plot  is allocated proportionally to its signal. For each ``data plot'' we create five sets of null plots to be evaluated simultaneously $(K_1, ..., K_5)$ times.  
$p$-values are then based on a comparison of the sum of data picks from  five no-signal lineups and  the observed number of data picks from the actual lineups. The column on the right of Table~\ref{tab:results} shows $p$-values based on $10^5$ simulation runs.

\begin{table}[ht]
\centering
\caption{\label{tab:results} Overview of all lineup evaluations. Ratios comparing the number correct to the total number of evaluations are shown.  $p$-values and significances are based on the calculations as described in Section~\ref{sec:pvalues}.
} 
\begin{tabular}{Mrrlrlrlrlrlc}
  \hline
&  & \multicolumn{9}{c}{Replicate} && Overall \\ \cline{3-12} 
\multicolumn{2}{c}{Lineup} & 1 & & 2 && 3 && 4 && 5 && $p$-values\\ 
  \hline
  
\includegraphics[width=0.05\textwidth]{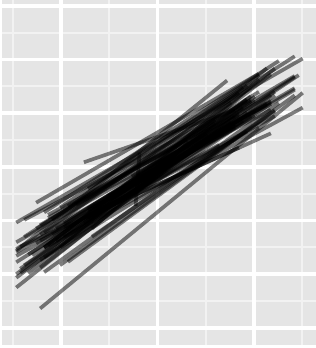}&  fig.~\ref{fig:fanned} &   10/68 & \hspace{-0.1in}* & 7/65 & \hspace{-0.1in}  & 8/61 & \hspace{-0.1in}. & 13/61 & \hspace{-0.1in}*** & 6/66 & \hspace{-0.1in}  & 0.0022 \\ 
 \includegraphics[width=0.05\textwidth]{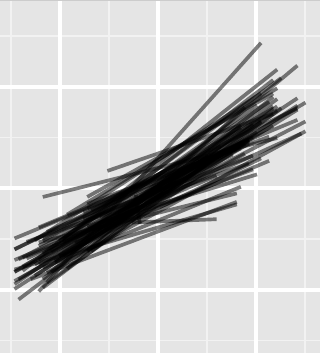} & fig.~\ref{fig:fanned2} & 0/64 & \hspace{-0.1in}  & 7/75 & \hspace{-0.1in}  & 11/76 & \hspace{-0.1in}* & 0/69 & \hspace{-0.1in}  & 0/60 & \hspace{-0.1in} & 0.4202 \\ 
\includegraphics[width=0.05\textwidth]{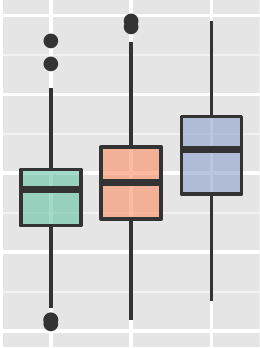}&   fig.~\ref{fig:boxplot-ordered} & 60/68 & \hspace{-0.1in}*** & 51/59 & \hspace{-0.1in}*** & 59/64 & \hspace{-0.1in}*** & 51/60 & \hspace{-0.1in}*** & 62/71 & \hspace{-0.1in}*** & $< 10^{-4}$ \\ 
 \includegraphics[width=0.05\textwidth]{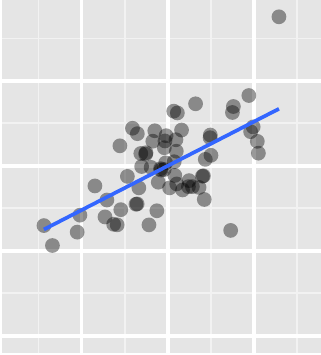}&    fig.~\ref{fig:ranef-corr} & 36/63 & \hspace{-0.1in}*** & 58/69 & \hspace{-0.1in}*** & 30/60 & \hspace{-0.1in}*** & 51/68 & \hspace{-0.1in}*** & 52/63 & \hspace{-0.1in}*** & $< 10^{-4}$ \\ 
\includegraphics[width=0.05\textwidth]{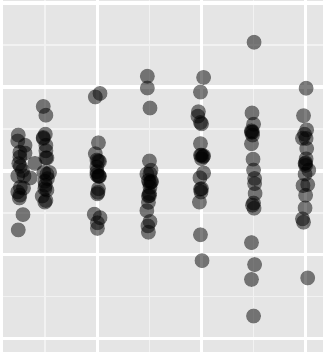}& fig.~\ref{fig:constvar2} & 26/80 & \hspace{-0.1in}*** & 9/59 & \hspace{-0.1in}* & 23/60 & \hspace{-0.1in}*** & 7/55 & \hspace{-0.1in}. & 11/69 & \hspace{-0.1in}* & $< 10^{-4}$ \\ 
\includegraphics[width=0.05\textwidth]{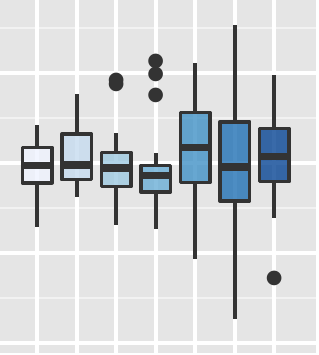} &  fig.~\ref{fig:constvar2.bp} & 23/70 & \hspace{-0.1in}*** & 9/74 & \hspace{-0.1in}. & 11/62 & \hspace{-0.1in}** & 31/78 & \hspace{-0.1in}*** & 25/61 & \hspace{-0.1in}*** & $<10^{-4}$\\ 
\includegraphics[width=0.05\textwidth]{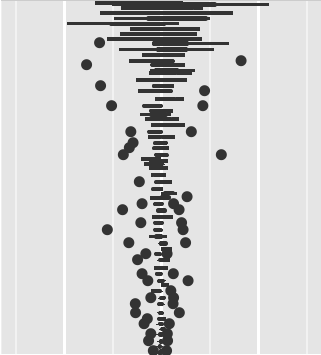}&   fig.~\ref{fig:badcyclone} & 49/73 & \hspace{-0.1in}*** & 41/59 & \hspace{-0.1in}*** & 41/59 & \hspace{-0.1in}*** & 40/66 & \hspace{-0.1in}*** & 49/65 & \hspace{-0.1in}*** & $< 10^{-4}$ \\ 
\includegraphics[width=0.05\textwidth]{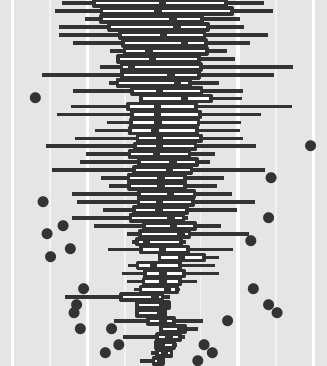} & fig.~\ref{fig:goodcyclone} &1/59 & \hspace{-0.1in}  & 2/79 & \hspace{-0.1in}  & 2/68 & \hspace{-0.1in}  & 4/62 & \hspace{-0.1in}  & 1/72 & \hspace{-0.1in}  & 0.6567 \\ 
\includegraphics[width=0.05\textwidth]{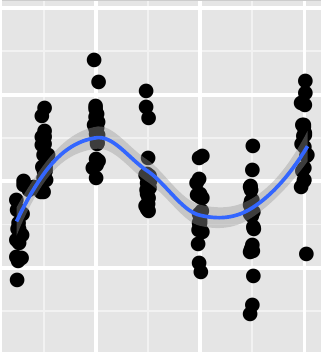}&   fig.~\ref{fig:linearity} & 52/55 & \hspace{-0.1in}*** & 60/62 & \hspace{-0.1in}*** & 49/52 & \hspace{-0.1in}*** & 79/83 & \hspace{-0.1in}*** & 63/67 & \hspace{-0.1in}*** & $< 10^{-4}$ \\ 
\includegraphics[width=0.05\textwidth]{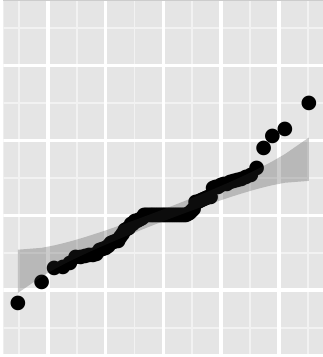}&   fig.~\ref{fig:qqlineup-t} & 26/63 & \hspace{-0.1in}*** & 48/75 & \hspace{-0.1in}*** & 0/59 & \hspace{-0.1in}  & 11/56 & \hspace{-0.1in}** & 6/69 & \hspace{-0.1in}  & $< 10^{-4}$ \\ 
\multicolumn{2}{r}{random intercept} & 0/72 & \hspace{-0.1in}  & 1/75 & \hspace{-0.1in}  & 0/68 & \hspace{-0.1in}  & 0/75 & \hspace{-0.1in}  & 2/61 & \hspace{-0.1in} & 0.8904 \\ 
\multicolumn{2}{r}{random slope} & 0/65 & \hspace{-0.1in}  & 0/64 & \hspace{-0.1in}  & 0/68 & \hspace{-0.1in}  & 0/46 & \hspace{-0.1in}  & 0/64 & \hspace{-0.1in}  & 1.0000\\ 
\includegraphics[width=0.05\textwidth]{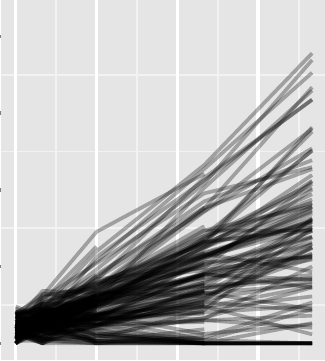}&   fig.~\ref{fig:autism-ranef} & 48/76 & \hspace{-0.1in}*** & 26/55 & \hspace{-0.1in}*** & 28/63 & \hspace{-0.1in}*** & 30/70 & \hspace{-0.1in}*** & 30/60 & \hspace{-0.1in}*** & $< 10^{-4}$ \\ 
 \includegraphics[width=0.05\textwidth]{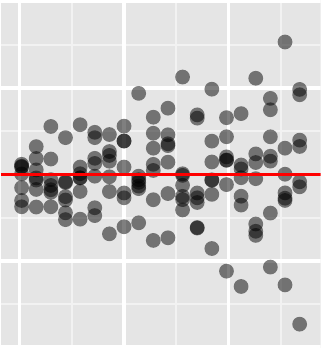} & fig.~\ref{homogeneous-1}&0/57 & \hspace{-0.1in}  & 1/71 & \hspace{-0.1in}  & 8/75 & \hspace{-0.1in}  & 0/59 & \hspace{-0.1in}  & 2/68 & \hspace{-0.1in}  & 0.6448 \\ 

\includegraphics[width=0.05\textwidth]{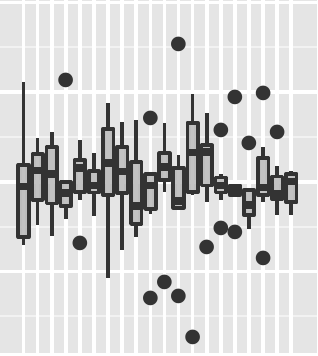} &  fig.~\ref{homogeneous-2}& 23/61 & \hspace{-0.1in}*** & 38/72 & \hspace{-0.1in}*** & 29/59 & \hspace{-0.1in}*** & 22/70 & \hspace{-0.1in}*** & 35/62 & \hspace{-0.1in}*** & $< 10^{-4}$ \\
   \hline
\multicolumn{12}{l}{Signif. codes:  0 $\le$ *** $\le$ 0.001 $\le$ ** $\le$ 0.01 $\le$ * $\le$ 0.05 $\le$ . $\le$ 0.1 $\le$ ' ' $\le$ 1}

\end{tabular}
\end{table}

\begin{table}[ht]
\caption{\label{tab:reasons} Percent of data picks,  given the reason for the choice of plot from the lineup. }
\centering
\begin{tabular}{Mrrrrrrr}
  \hline
\multicolumn{2}{c}{Lineup} & Outlier & Spread & Trend & Asymmetry & Other \\ 
  \hline
\includegraphics[width=0.05\textwidth]{examfanned-icon} &  fig.~2  & 12.8 & 24.0 & 10.9 & 5.8 & 15.9 \\ 
 \includegraphics[width=0.05\textwidth]{exam-with-slope-icon} & fig.~4 & 2.4 & 10.6 & 3.5 & 5.4 & 0.0 \\ 

\includegraphics[width=0.05\textwidth]{autism-ordered-icon} &   fig.~3 & 73.3 & 95.5 & 92.4 & 91.0 & 80.2 \\ 

 \includegraphics[width=0.05\textwidth]{examcorr-icon}&   fig.~5 & 64.5 & 34.3 & 83.4 & 70.7 & 64.9 \\ 

\includegraphics[width=0.05\textwidth]{dialyzerheterogeneous-icon}& fig.~6 & 19.0 & 27.3 & 29.6 & 25.8 & 19.0 \\ 
\includegraphics[width=0.05\textwidth]{dialyzerheterogeneous-bp}& fig.~\ref{fig:constvar2.bp} & 26.7 & 24.6 & 30.1 & 43.7 & 0.0 \\

\includegraphics[width=0.05\textwidth]{cyclone-icon}&   fig.~7 & 49.6 & 49.5 & 77.7 & 79.4 & 91.8 \\ 
\includegraphics[width=0.05\textwidth]{cyclone-good-icon}&   fig.~8  & 4.2 & 0.6 & 2.4 & 6.8 & 0.0 \\ 

\includegraphics[width=0.05\textwidth]{dialyzernonlinear-icon}&   fig.~9 & 84.0 & 87.4 & 98.3 & 98.1 & 100.0 \\ 

\includegraphics[width=0.05\textwidth]{radontranef-icon}&   fig.~11 & 37.8 & 49.2 & 19.6 & 4.4 & 10.9 \\ 
\multicolumn{2}{r}{random intercept} & 1.5 & 0.0 & 1.1 & 0.0 & 0.0 \\ 
\multicolumn{2}{r}{random slope} & 0.0 & 0.0 & 0.0 & 0.0 & 0.0 \\ 
\includegraphics[width=0.05\textwidth]{autism2-fanned-icon}&   fig.~\ref{fig:autism-ranef}  & 39.2 & 33.8 & 57.9 & 78.5 & 70.0 \\ 
 \includegraphics[width=0.05\textwidth]{homogeneous-dots-icon} & fig.~\ref{homogeneous-1} & 2.4 & 3.0 & 7.6 & 4.8 & 0.0 \\ 
 \includegraphics[width=0.05\textwidth]{homogeneous-bp-icon} & fig.~\ref{homogeneous-2} & 56.8 & 55.1 & 20.1 & 33.9 & 22.7 \\ 
   \hline
\end{tabular}
\end{table}
\clearpage
\subsection{Generating null plots}\label{app:nullplots}

All of the lineups presented in this paper use a parametric bootstrap to generate plots consistent with the null hypothesis. This section outlines the parametric bootstrap for LME models and provides more detail about how the null plots were generated.

For a fitted continuous response LME model (as outlined by Equations \eqref{eq:hlm}--\eqref{eq:conditionalmod} and the intermediate discussion) the parametric bootstrap proceeds as follows:

\begin{enumerate}
\item Generate a vector of $q$ random effects (i.e., level-2 residuals) from $\mathcal{N}(\bm{0},\ \widehat{\bm{D}})$ for each group; that is, generate $\bm{b}^*_i \sim \mathcal{N}(\bm{0},\ \widehat{\bm{D}})$ for $i = 1,\ldots, g$. 
\item Generate a vector of residuals of length $n_i$ (i.e., level-1 residuals) from $\mathcal{N}(\bm{0},\ \widehat{\sigma}^2 \widehat{\bm{R}_i})$ for each group; that is, generate $\bm{\varepsilon}^*_i \sim \mathcal{N}(\bm{0},\ \widehat{\sigma}^2 \widehat{\bm{R}_i})$ for $i = 1,\ldots, g$.
\item Generate a bootstrap sample $\bm{y}_i^*$ for each group $i=1,\ldots,g$ from $\bm{y}_i^* = \bm{X}_i \widehat{\bm{\beta}} + \bm{Z}_i \bm{b}_i^* + \bm{\varepsilon}^*_i$.
\item Refit the model to the bootstrap samples.
\item Repeat steps 1--4 $B$ times.
\end{enumerate}

The parametric bootstrap was used to generate the null plots in each situation. 
During model selection the simpler models are used to generate the null plots using the parametric bootstrap, while model checking bootstraps the original model to generate null plots.

\paragraph{Selecting fixed effects.} To use lineup tests to determine whether a variable should be included in the fitted model we must generate null plots from a model \emph{excluding} the variable in question. Let $\bm{X}_i^{(c)}$ denote the design matrix for the fixed effects with the $c$th column  deleted. We use the model 
\[
\bm{y}_i = \bm{X}_i^{(c)} \bm{\beta} + \bm{Z}_i \bm{b}_i + \bm{\varepsilon}_i
\]
to generate the null plots.

\paragraph{Selecting random effects structure.} To use a lineup test to determine whether a random effect should be included in the fitted model we must generate null plots from a model \emph{excluding} the random effect in question. This results in a model of the form of~\eqref{eq:hlm} with the column of $\bm{Z}_i$ corresponding to the random effect in question deleted, and a random effects vector $\bm{b}_i$ of length $q-1$. To determine whether it is necessary to allow the random effects to be correlated, the null plots are generated using a model where the covariance matrix of the random effects $\bm{D}$ has zero entries in the appropriate off diagonal entries.

\paragraph{Model checking.} Generating the null plots used for model checking is a direct application of the parametric bootstrap as detailed above to obtain $B$ fitted models from which the appropriate aspects are extracted.

\subsection{Additional lineups included in the study}\label{app:morelineups}

This section includes two lineups that were included in the  MTurk study, but were not discussed in the paper. 

Figure~\ref{fig:constvar2.bp} contains a box plot representation of the same data as Figure~5 in the paper, but categorizes pressure into seven categories, and shows residuals in the form of box plots. In order to preserve the appearance of continuity on the $x$-axis we used a color scheme to fill the boxes with deepening shades of blue from left to right. In this form 23 out of 70 observers identify the plot of the data.  This is consistent with the other design.

\begin{figure}[hbt]
	\centering
	\includegraphics[width=0.8\textwidth]{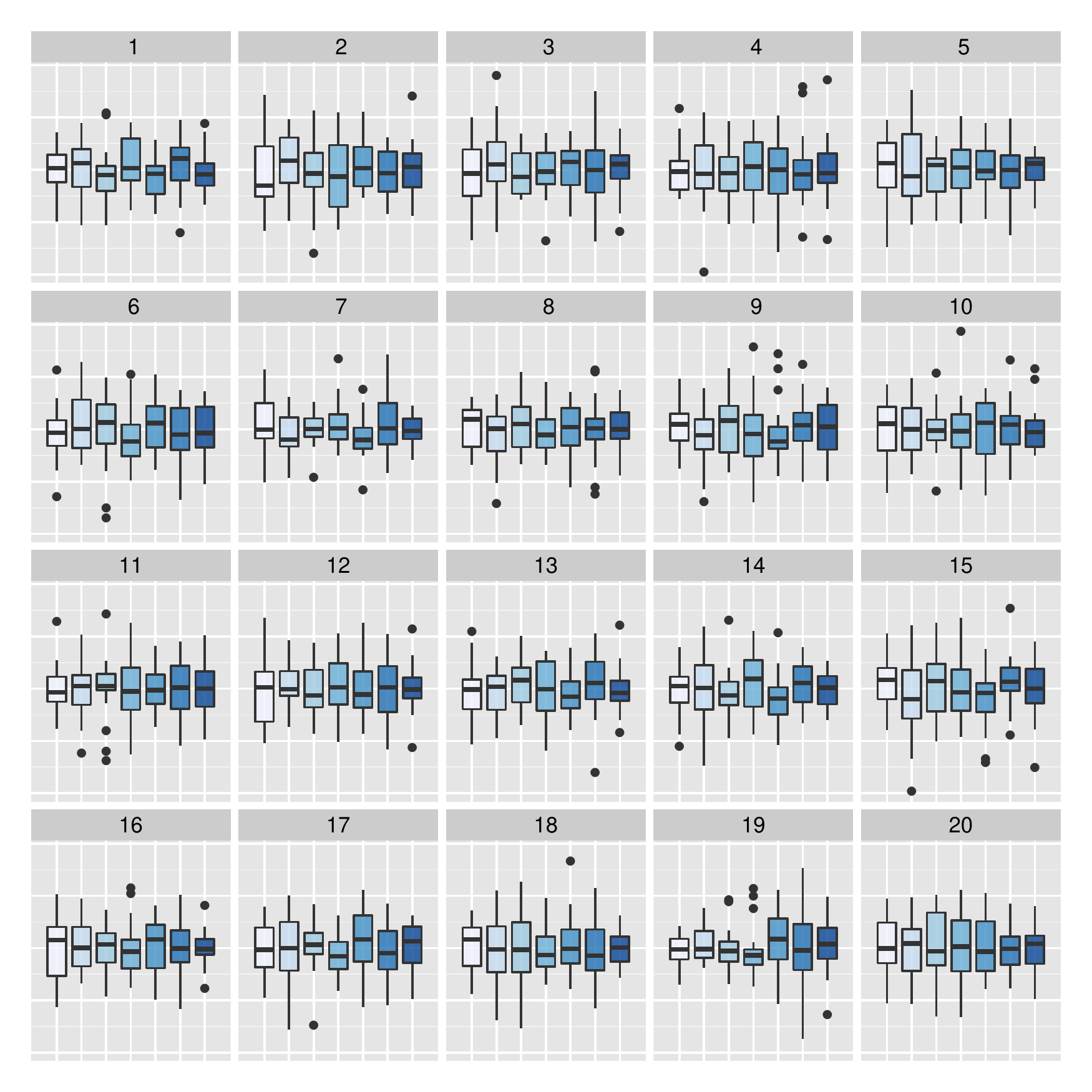}
	\caption{\label{fig:constvar2.bp} 
	Alternative box plot  Lineup testing homogeneity of the level-1 residuals. Which of the plots is the most different? Which feature led you to your choice?}
\end{figure}


Figure~\ref{fig:autism-ranef} displays another lineup testing the adequacy of the random effects specification  (see Section~3.2 of the paper) using data from the autism study. The null plots were generated from a model containing only a linear random slope, so if the true plot in panel \#($\sqrt{16} + 12$) is identified it provides support for the inadequacy of this specification, and the need for additional random effects. 

\begin{figure}
	\centering
	\includegraphics[width=0.8\textwidth]{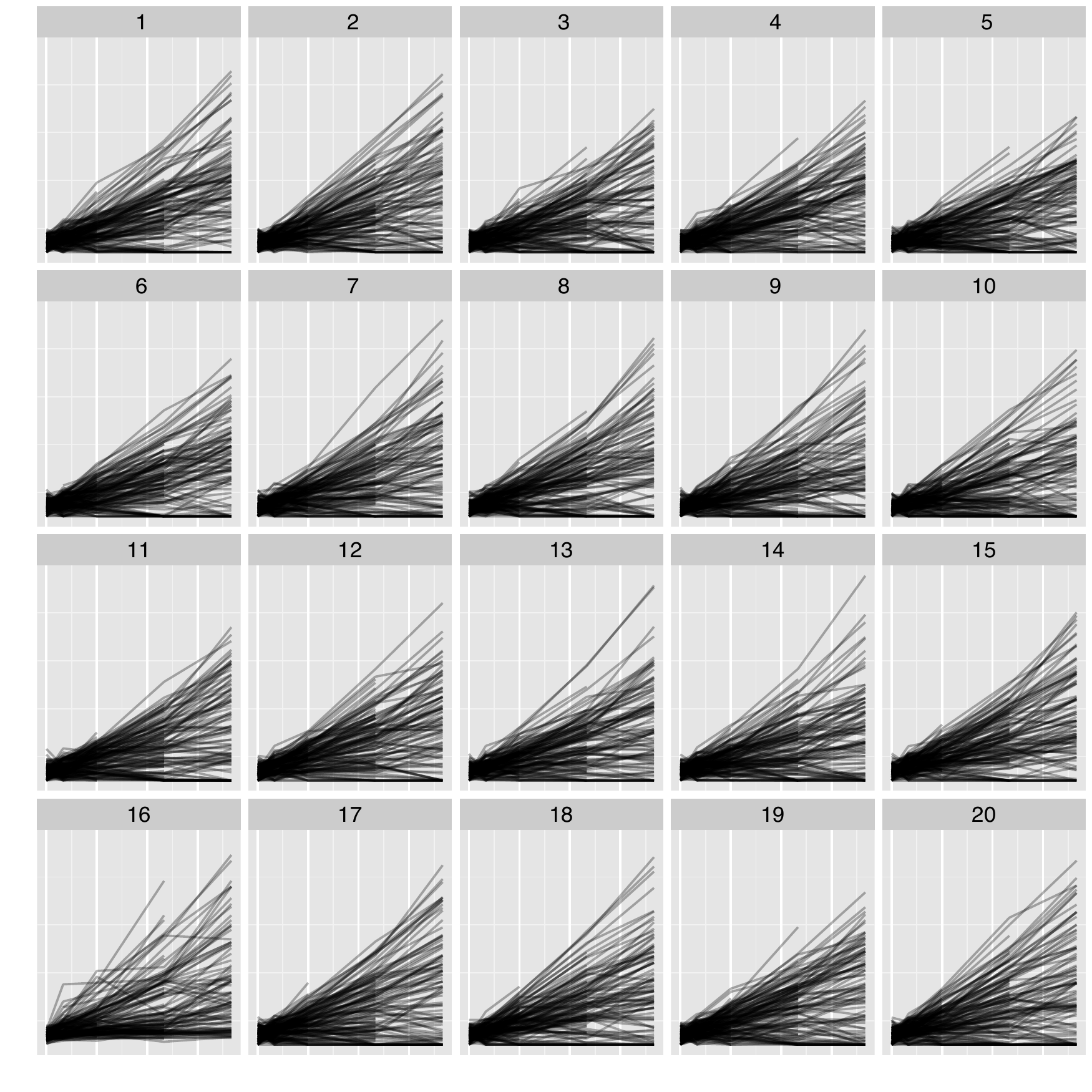}
	\caption{\label{fig:autism-ranef} Which of the plots is the most different? Which feature led you to your choice? }
\end{figure}



\begin{figure}[hbt]
	\centering
	\includegraphics[width=0.8\textwidth]{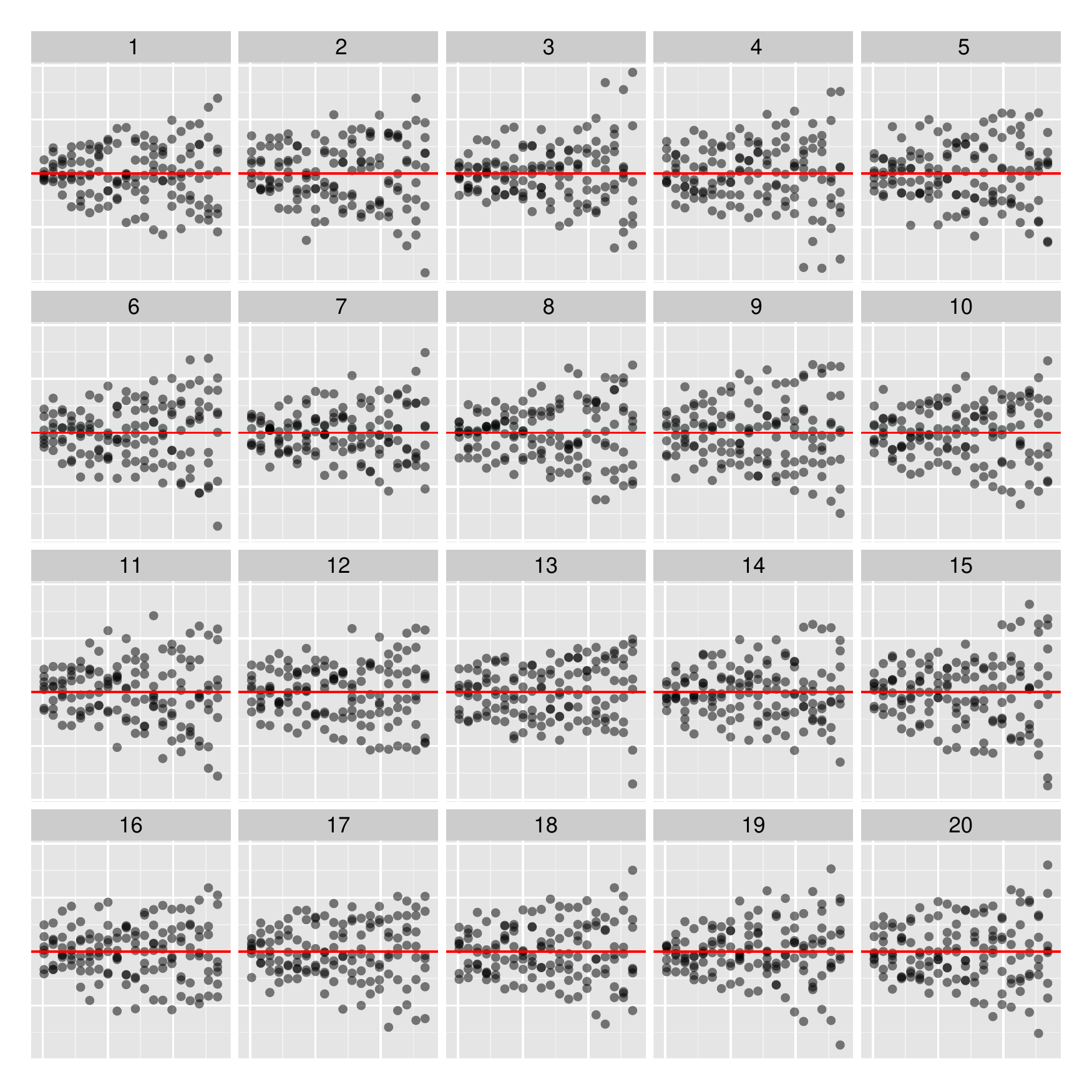}
	\caption{\label{homogeneous-1}
	Lineup testing homogeneity of the level-1 residuals. Which of the plots is the most different? Which feature led you to your choice?}
\end{figure}
\begin{figure}[hbt]
	\centering
	\includegraphics[width=0.8\textwidth]{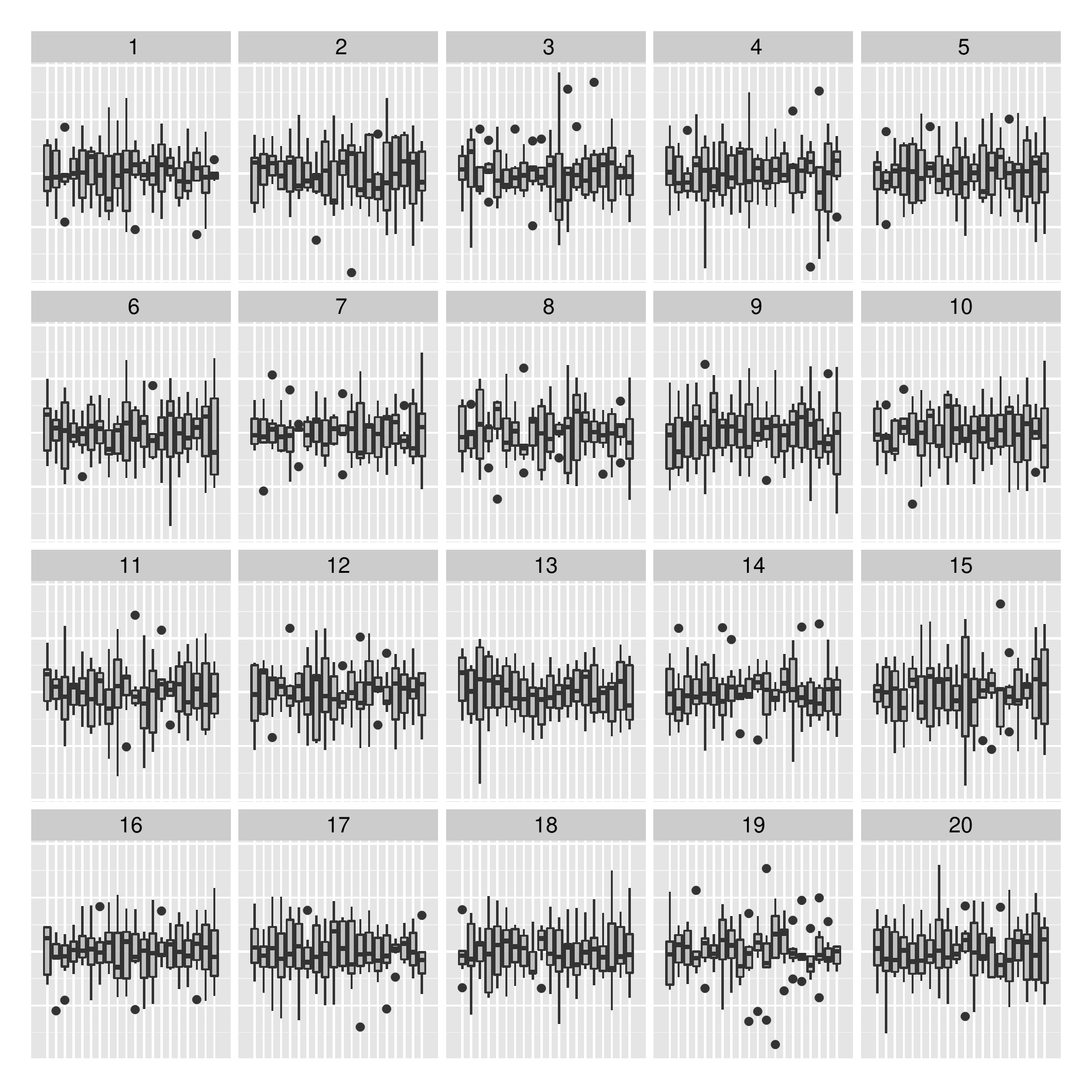}
	\caption{ \label{homogeneous-2}
	Lineup testing homogeneity of the level-1 residuals. Which of the plots is the most different? Which feature led you to your choice?}
\end{figure}


Figures~\ref{homogeneous-1} and~\ref{homogeneous-2} show another example of testing for homogeneity in the variance following the approach taken in Section~4.1 of the paper. Both of these lineups are based on the dialyzer data. 
Level-1 residuals are plotted by subject. Subjects are ordered by variance---i.e.,~we get some structure that might be taken for differences in variability, that are really just differences due to the imbalance in group size. If any panel of this lineup is considered separately, an analyst may come to the conclusion that the within-group variance increases across the $x$ axis.  However, inserting the true plot into the lineup forces the analyst to consider this particular feature as inherent to the data structure rather than evidence against a hypothesis of homogenous variance. 
The dot plot version is not significant, but the box plot version is. When participants in the box plot design identify the data plot, about 45.3\% give outliers as the reason for their choice. In contrast to that, outliers, a large spread or a trend are in a three-way tie for the reason for identifying the data plot in the lineup with the dot plot design.

The reason for the box plot design being so much more significant might not be so much of an issue of homogeneity being violated as much as a difference in the error distribution between the data and the nulls. Null data come from a parametric bootstrap where residuals are simulated under a normal error assumption. Outliers in small samples are indicative of the sample being from a distribution with heavy tails.
Regardless of the reasoning, the second lineup design enables us to diagnose a problem with the model that makes the data stand out from a set of nulls.  The two designs therefore represent two very similar tests with different power.

\clearpage

\bibliographystyle{apalike}
\bibliography{hlmviz_bib}

\end{appendix}

\end{document}